\documentclass[11pt, a4paper]{article}
\usepackage[english]{babel}
\usepackage[utf8]{inputenc}
\usepackage{amsmath,bm}
\usepackage{amssymb}
\usepackage{tabularx} 
\usepackage{threeparttable}
\usepackage{ragged2e}
\usepackage{multirow}
\usepackage{booktabs}
\usepackage{xcolor}
\usepackage{placeins}
\usepackage{soul}
\usepackage{algorithm}
\usepackage{algpseudocode}
\usepackage{appendix}
\usepackage[margin=2.5cm]{geometry}
\usepackage{graphicx}
\usepackage[colorlinks,citecolor=blue,urlcolor=blue, linkcolor=blue,bookmarks=false,hypertexnames=true]{hyperref}

\usepackage{csquotes}
\usepackage[natbib=true,
            style=authoryear,
            maxbibnames=10,
            uniquename=false]{biblatex}

\AtEveryBibitem{
    \clearfield{issn}
    \clearfield{eprint}
    \clearfield{urlyear}
    \clearfield{month}
    }

\DeclareFieldFormat{sentencecase}{\MakeSentenceCase{#1}}

\renewbibmacro*{title}{%
  \ifthenelse{\iffieldundef{title}\AND\iffieldundef{subtitle}}
    {}
    {\ifthenelse{\ifentrytype{article}\OR\ifentrytype{inbook}%
      \OR\ifentrytype{incollection}\OR\ifentrytype{inproceedings}%
      \OR\ifentrytype{inreference}}
      {\printtext[title]{%
        \printfield[sentencecase]{title}%
        \setunit{\subtitlepunct}%
        \printfield[sentencecase]{subtitle}}}%
      {\printtext[title]{%
        \printfield[titlecase]{title}%
        \setunit{\subtitlepunct}%
        \printfield[titlecase]{subtitle}}}%
     \newunit}%
  \printfield{titleaddon}}

\newcommand{\thickdot}{\mathbin{\vcenter{\hbox{\scalebox{1.5}{$\cdot$}}}}}

\makeatletter
\newcommand{\multiline}[1]{%
  \begin{tabularx}{\dimexpr\linewidth-\ALG@thistlm}[t]{@{}X@{}}
    #1
  \end{tabularx}
}
\makeatother

\newcommand{\Gammad}[1]{ \mathcal{G}\left(#1\right)}

\begin{document}

\title{Bayesian nonparametric partial clustering: Quantifying the effectiveness of agricultural subsidies across Europe}

\author{Alexander Mozdzen$^{a}$\thanks{Corresponding author at: Department of Statistics, University of Klagenfurt, Universit\"atsstra\ss{}e 65-67, 9020 Klagenfurt am W\"orthersee, Austria. Email: alexander.mozdzen@aau.at} , Felicity Addo$^{b}$, Tamas Krisztin$^{b}$, Gregor Kastner$^{a}$\\
        \small $^{a}$Department of Statistics, University of Klagenfurt, Klagenfurt, Austria \\
        \small $^{b}$ International Institute for Applied Systems Analysis (IIASA), Laxenburg, Austria \\
}

\renewcommand{\thefootnote}{}%
\footnotetext{Alexander Mozdzen and Gregor Kastner acknowledge funding from the Austrian Science Fund (FWF) for the project ``High-dimensional statistical learning: New methods to advance economic and sustainability policy'' (ZK 35), jointly carried out by the University of Klagenfurt, Paris Lodron University Salzburg, TU Wien, and the Austrian Institute of Economic Research (WIFO).

Felicity Addo and Tamas Krisztin acknowledge funding from the European Union’s Horizon Europe Research and Innovation Programme under Grant Agreement No. 101060423, as well as from the Jubilee Fund of the Austrian National Bank (OeNB), Project No. 18307.}
 
\maketitle

\begin{abstract} 
\noindent 
The global climate has underscored the need for effective policies to reduce greenhouse gas emissions from all sources, including those resulting from agricultural expansion, which is regulated by the Common Agricultural Policy (CAP) across the European Union (EU). To assess the effectiveness of these mitigation policies, statistical methods must account for the heterogeneous impact of policies across different countries. We propose a Bayesian approach that combines the multinomial logit model, which is suitable for compositional land-use data, with a Bayesian nonparametric (BNP) prior to cluster regions with similar policy impacts. To simultaneously control for other relevant factors, we distinguish between cluster-specific and global covariates, coining this approach the Bayesian nonparametric partial clustering model. We develop a novel and efficient Markov Chain Monte Carlo (MCMC) algorithm, leveraging recent advances in the Bayesian literature. Using economic, geographic, and subsidy-related data from 22 EU member states, we examine the  effectiveness of policies influencing land-use decisions across Europe and highlight the diversity of the problem. Our results indicate that the impact of CAP varies widely across the EU, emphasizing the need for subsidies to be tailored to optimize their effectiveness.
\end{abstract}


\section{Introduction}

Land-use change is a significant driver of climate change, biodiversity loss, and ecosystem degradation across Europe \citep{burrascano2016current, kuemmerle2016hotspots}. The conversion of natural habitats, such as forests and grasslands, into intensive agricultural land disrupts carbon sinks, fragments ecosystems, and diminishes essential ecosystem services \citep{plieninger2016driving}, posing a critical threat to biodiversity \citep{hulber2017plant, reidsma2006impacts}. To mitigate these environmental challenges and achieve Europe’s climate change commitments, it is essential to design policies that balance conservation efforts with the sustainable use of land.

In the European Union (EU), the Common Agricultural Policy (CAP) serves as the primary regulatory framework governing agricultural practices and land-use patterns. Recent CAP reforms aim to harmonize agricultural productivity with environmental sustainability and rural development goals, addressing evolving social and environmental challenges. By offering income support, production incentives, and environmental subsidies, the CAP significantly shapes land-use decisions, influencing how land is allocated between croplands, grasslands, forests, and other natural vegetation. Numerous studies have demonstrated the CAP's substantial role in driving land-use dynamics \citep{renwick2013policy, helming2018economic, peer2019greener} and shaping land-use changes in different regions of Europe \citep{kuemmerle2016hotspots, cortignani2019cap,  krisztin2022spatial}.

However, the effects of the CAP on land use are highly heterogeneous across Europe, varying with the diversity of agricultural systems, environmental conditions, and socio-economic contexts. A policy that successfully supports sustainable land use in one region may prove less effective or even counterproductive in another. Recent research highlights this spatial heterogeneity, linking regional land-use changes to factors such as climate variability and farm management practices \citep{ reidsma2006impacts,kuemmerle2016hotspots,levers2018spatial}. The economic impacts of CAP subsidies also vary, with regional disparities often magnified by climate-related challenges \citep{giannakis2015highly}. For instance, \citet{rega2019environmentalism, overmars2013modelling} illustrate how policy choices may prioritize either agricultural productivity or biodiversity conservation, depending on the region.

This variation underscores the compositional nature of land-use decisions, where agricultural subsidies often favor the expansion of arable land, potentially at the expense of other land-use types, leading to environmental trade-offs. For example, \citet{oneill2020forest} explores the decision-making process between maintaining land for sheep farming and converting it to forests for carbon sequestration. \citet{helming2018economic} argue for carefully structured subsidy schemes that balance competing land-use interests to minimize adverse environmental impacts.

Given the compositional structure of land-use data, multinomial logit (MNL) models are a natural choice for statistical analysis, with various applications in the literature \citep[see, for example,][]{debellagilo_2009_norway_mnl, temme2011mapping, hao2015integration, krisztin2022spatial}.

However, the spatial nature of land-use dynamics necessitates models that account for spatial dependencies. For instance, \citet{krisztin2022spatial} incorporated spatial dependence through a spatial autoregressive (SAR) term using an exogenous neighborhood matrix. While this model estimates the strength of spatial autocorrelation for each land-use class, it lacks variability as all areal units share the same regression coefficients. In contrast, \citet{temme2011mapping} proposed a model with separate regression coefficients for different regions, estimating a distinct multinomial logit model for each region, thus excluding the sharing of information across regions.

To bridge the gap between the approaches of \citet{krisztin2022spatial} and \citet{temme2011mapping}, we propose an infinite mixture of multinomial logit models, clustering groups of areal units based on their regression coefficients. Our model distinguishes between covariates that influence clustering, having a separate impact in each cluster, and those with a global impact. This approach has several advantages: enabling partial pooling through clustering, providing a spatial mapping based solely on chosen covariates without relying on exogenous information, and controlling for homogeneous effects through partial clustering. 

In our empirical application, we choose seven subsidy-related \textit{cluster-covariates} and six \textit{global covariates} for the analysis. While the former directly impact the clustering through their estimated regression coefficients, the latter serve as control variables with a global impact, thereby leading to the designation \textit{partial clustering}. This choice is well motivated by the question at hand, aiming to uncover the cluster-specific impact of farm subsidies across Europe while simultaneously considering other relevant variables like employment, GDP, or the respective elevation and slope.

Our model incorporates spatial variability and recognizes the interconnectedness of land-use changes, offering a nuanced understanding of how CAP subsidies promoting one type of land use may adversely affect others, with these effects varying significantly across different regional contexts. A similar approach was recently proposed by \citet{papastamoulis_2023_model_clustering}, who introduced software for estimating finite mixtures of multinomial logit models. While \citet{papastamoulis_2023_model_clustering} employs finite mixtures and Metropolis-Hastings steps for estimating regression coefficients, we opt for the infinite mixture model and introduce a novel Gibbs-type scheme that relies entirely on samples from full conditional distributions. Additionally, we introduce the concept of partial clustering, which, to our knowledge, has not yet been explored in the literature.

Our empirical analysis covers 21 countries and their respective NUTS-3 regions over the period 2007 to 2017, focusing on four primary land-use categories: arable land, grassland, forest, and urban areas. Land that does not fit into these categories is classified as ``other natural vegetation''. Using farm-level data from the Farm Accountancy Data Network (FADN), we evaluate both Pillar I and Pillar II CAP policies across various geographic contexts. This comprehensive dataset facilitates a detailed assessment of the regional effectiveness of the CAP and its diverse impacts on land-use patterns.

The remainder of the paper is organized as follows: Section~\ref{sec:meth} details the methodological approach, including BNP clustering and partial clustering in the context of the multinomial logit model. The novel MCMC algorithm and necessary derivations are presented in Section~\ref{sec:bayes_inf}, before describing the dataset in Section~\ref{sec:data_2}. The empirical results are discussed in Section~\ref{sec:results}, and the paper concludes with recommendations in Section~\ref{sec:conclusion_2}.


\section{Methodology}
\label{sec:meth}

We estimate an econometric model that aims at explaining the choice of land use in $N$ NUTS-3 regions across Europe. In a given region $i$ (with $i = 1, \dots , N$), at each time $t$ (with $t = 1, \dots, T$) land owners may utilize their land for $J$ distinct land uses. In our case these are cropland, grassland, forest, urban, and other natural vegetation. 

For easier notation, we encode all observations in the three-dimensional array $ \bm{Y} \in \mathbb{R}^{N \times T \times J} $. We denote with $\thickdot$ all entries of the corresponding dimension, such that $\bm{Y}_{i t \thickdot} = (y_{i t 1}, \dots, y_{i t J})^{T} $ describes a column vector including observations for areal unit $i$ at time point $t$ for all categories. Accordingly, using two $\thickdot$ in the index denotes a matrix where the row dimension corresponds to the variable replaced by the first $\thickdot$ and the column dimension corresponds to the variable replaced by the second $\thickdot$.


In what is to follow, we start by giving an introduction to Bayesian nonparametric (BNP) clustering and Dirichlet process mixtures in Section~\ref{sec:bnp}. Section~\ref{sec:bnppc} lays out how these ideas can be extended to allow for Bayesian nonparametric \emph{partial} clustering (BNP-PC). Finally, Section~\ref{sec:multinom}, applies the proposed framework to the multinomial logistic regression model.

\subsection{Bayesian nonparametric (BNP) clustering} \label{sec:bnp}
Defining an individual set of regression coefficients $\bm{B}_{i \thickdot \thickdot}$ for every areal unit, which in our application would amount to $N \times (K + 1) \times (J - 1) = 51072$ parameters, can quickly lead to an unfavorable amount of estimable parameters, challenging to estimate and interpret and prone to overfitting. Conversely, complete pooling of information by estimating one set of parameters for all areal units, i.e. keeping $\bm{B}_{i \thickdot \thickdot}$ constant for all $i$, thus reducing the number of parameters by the factor $N$, might lead to an oversimplified representation, lacking in comprehensive detail. We aim to strike a balance between these alternatives and adhere to Occam's razor by clustering the areal units in an automatic and model-based fashion \citep[cf., e.g.,][]{fraley_1998_howmanyclust, chandra_2023_curseodim_mdbclust}.

Clustering multinomial logit models using finite mixtures has been explored in the Bayesian framework via MCMC sampling in the works of \citet{fruehwirth_schnatter_logreg_mix} and in the frequentist setting via expectation maximization (EM) in \citet{gruen_2008_mulnom_mix} or \citet{benaglia_2009_mixtools}. In this work we opt for the nonparametric counterpart of finite mixtures using the popular Dirichlet process (DP), so-called Dirichlet Process mixtures (DPM). We adopt this method owing to its computationally convenient representations \citep{blackwell_1973_polya_urn, sethuraman_94_stick_break}, which have led to the development of efficient sampling mechanisms, most notably the MCMC algorithms introduced in \citet{Lo_1984_dpms, west_1995_hyperparameter, Neal00, maceachern_1998_nogaps} and the variational inference algorithm by \citet{blei_2006_vi_dpm}. Since \citet{mfm} showed that many of these favorable properties can be translated into the framework of mixtures of finite mixtures, we do not delve into the debate about the finite or infinite choice \citep{miller_2014_inconsistency, giordano_2023_bnp_sensitivity, fruehwirth_schnatter_2019_heretoinfinity, ascolani_2023_clust_consistency}, but refine our focus on the novel partial clustering method, efficient MCMC scheme as well as empirical contribution. \citet{mozdzen_2022_bstc} use a similar approach of using a DP to cluster regression coefficients, but do so in the context of a spatio-temporal linear regression setting.

The DPM can be expressed in its hierarchical form as 
\begin{align*} 
 \bm{Y}_{i \thickdot \thickdot}\mid 
\bm{X}_{i \thickdot \thickdot}, \bm{B}_{i \thickdot \thickdot} &\stackrel{\text{ind}}{\sim} 
f\left(\bm{Y}_{i \thickdot \thickdot} \mid 
\bm{X}_{i \thickdot \thickdot},  \bm{B}_{i \thickdot \thickdot}\right), 
\quad i=1 \ldots N, \\
 \bm{B}_{i \thickdot \thickdot} \mid G &\stackrel{\mathrm{iid}}{\sim} G, \\
 G &\sim DP\left(\alpha, G_0 \right),\\
\alpha &\sim \Gammad{\alpha_0, \beta_0},
\end{align*}
where $f(\thickdot )$ is a suitable kernel, in our case the kernel of a multinomial distribution and $G$ a sample from the DP with concentration parameter $\alpha$ and base measure $G_0$. 

A sample $G \sim DP(\alpha, G_0)$ is a discrete probability distribution that exhibits a stronger resemblance to $G_0$ the larger $\alpha$ is \citep{ferguson_1983_dpm}. Due to this discreetness, multiple areal units are assigned the same set of coefficients, which we then designate as belonging to the same cluster. To encode the clustering of the areal units, we introduce the vector $\bm{s} = (s_1, \dots, s_N)$, whose $i$th entry denotes the cluster allocation of the $i$th areal unit and can take values in $1, \dots, M$, where $M$ denotes the total number of clusters.
We follow \citet{ascolani_2023_clust_consistency}, who show that the infamous inconsistency of DPMs regarding the number of clusters \citep{miller_2014_inconsistency} can be mitigated by learning $\alpha$ in a fully Bayesian way. For this reason, we put a prior on $\alpha$, opting for a gamma distribution to ensure positive values and computational convenience.

As mentioned before, we are interested in the DP due to its advantageous representations regarding sampling algorithms. One of them is the P\'{o}lya urn scheme, which specifies the prior $p(\bm{s} \mid \alpha)$ as
\begin{align}\label{eq:s_prior}
&p(\bm s \mid \alpha) = P(s_1 \mid \alpha )\prod_{i = 2}^N P\left(s_i \mid s_1, \ldots, s_{i-1}, \alpha \right), \\
& P\left(s_i = c  \mid s_1, \ldots, s_{i-1}\right) = 
\begin{cases}
	&\dfrac{n_{ic}}{i-1 + \alpha}, \quad \text {for } 1 \leq c \leq M\\
	&\\
	&\dfrac{\alpha}{i-1 + \alpha}, \quad \text {for } c = M + 1
\end{cases} \nonumber,
\end{align}
where $n_{ic}$ is the number of units assigned to the $c$th cluster before the $i$th observation is considered. After the first value $s_1$ is standardly assigned to the first cluster, the following observations can either join existing clusters with probability proportional to their current size $n_{ic}$, or form a new cluster with probability proportional to $\alpha$, aptly highlighting the parameter's influence on the clustering behavior. 

While the quantification of uncertainty using the estimated posterior distribution is a sought-after strength of Bayesian methods, the analysis of the posterior clustering poses a challenge that requires more elaborate summarization methods than standard credible intervals or posterior modes. An appealing approach to finding a single partition representative of the posterior is via loss functions, out of which the Binder \citep{binder_1978_binder}, and Variation of Information \citep{vi} are two of the most prominent ones used for clustering problems. We refer the interested reader to the extensive overview about Bayesian clustering methods in \citet{wade_2023_bayclust_analysis}. After choosing a loss function, the optimal partition is obtained by minimizing the posterior expected loss. To this end, we use the stochastic search algorithm proposed in \citet{dahl_2022_salso_paper} and implemented in the R package \texttt{salso} \citep{dahl_2022_salso_package}. 
\subsection{Bayesian nonparametric partial clustering (BNP-PC)}\label{sec:bnppc}
With the inclusion of the DP prior, our model enables parsimonious, yet differentiated inference on clusters among the NUTS-3 units. Those regions, whose covariates have a similar estimated effect on the respective category of land use, are grouped together in the same cluster. A potential drawback of this approach is the implicit assumption that the impact of all covariates varies between clusters, thereby hindering the estimation of homogeneous effects.

To ameliorate this shortcoming and improve the flexibility of our model, we further augment it by splitting the data into variables we wish to cluster, $\bm{X}^{c} \in \mathbb{R}^{N \times T \times (K^{c} + 1)}$, and those we do not, $\bm{X}^{nc} \in \mathbb{R}^{N \times T \times K^{nc}}$, and introduce their corresponding sets of coefficients, $\bm{\beta} \in \mathbb{R}^{N \times (K^{c} + 1)  \times J}$ and $\bm{\theta}\in \mathbb{R}^{K^{nc}\times J}$, where $K = K^c + K^{nc}$. We refer to $X^c$ as cluster-covariates and $X^{nc}$ as global-covariates and to this model specification as Bayesian nonparametric-partial clustering (BNP-PC) model, whose hierarchical form reads as
\begin{align*} 
\bm{Y}_{i \thickdot \thickdot} \mid 
\bm{X}^{c}_{i \thickdot \thickdot}, \bm{X}^{nc}_{i \thickdot \thickdot},
\bm{\beta}_{i \thickdot \thickdot}, \bm{\theta}
&\stackrel{\text{ind}}{\sim} 
f\left(\bm{Y}_{i \thickdot \thickdot} \mid 
\bm{X}^{c}_{i \thickdot \thickdot}, \bm{X}^{nc}_{i \thickdot \thickdot},
\bm{\beta}_{i \thickdot \thickdot}, \bm{\theta} \right),
\quad i=1, \ldots, N, \\
\bm{\beta}_{i \thickdot \thickdot} \mid G 
& \stackrel{\mathrm{iid}}{\sim} G, \\
\bm{\theta} \mid F_0 
&\stackrel{\mathrm{iid}}{\sim} F_0, \\
G \mid \alpha &\sim DP\left(\alpha, G_0 \right),\\
\alpha &\sim \Gammad{\alpha_0, \beta_0},
\end{align*}
where $F_0$ is a suitable prior for the homogenous coefficients. We denote by $\bm{\beta}^{\star} \in \mathbb{R}^{M \times K^c \times J}$ the unique sets of parameters among all $\bm{\beta}$, such that $\bm{\beta}^{\star}_{c \thickdot \thickdot}$ includes all $\bm{\beta}_{i \thickdot \thickdot}$ that belong to cluster c. Section~\ref{sec:data_2} outlines the clustered and non-clustered covariates and gives detailed explanations motivating our choice.

Natural extensions of the BNP-PC model would be the explicit modeling of the spatial interrelations between the areal units or the temporal variation over the years. Both can be achieved through the inclusion of a so-called neighborhood matrix in a spatial- or conditional autoregressive term, respectively \citep{krisztin2022spatial, mozdzen_2022_bstc}, or simply by including the spatio-temporal information, e.g., as years and coordinates, in the set of explanatory variables \citep{augustin_2001_spatiotemporal_logit, debellagilo_2009_norway_mnl}. 
In this study, we choose not to include any exogenous spatial information, in order to learn the clustering and thereby the underlying spatial structure purely from the impact of $\bm{X}^{c}$ on the response variable. For the application at hand, this specification enables defining clusters of regions based on the impact of farm subsidies on land use, while simultaneously controlling for other economic covariates globally.
Note that in our case we choose to cluster across the dimension of covariates as well as categories, a model choice motivated by our research topic at hand. Interesting alternatives would be to cluster the areal units either based on covariates or categories, or to include two separate DPs and estimate two separate clusterings.

\subsection{The BNP-PC model in the context of multinomial logistic regression} 
\label{sec:multinom}
We showcase our methodological contribution using the multinomial logit regression model, which is suitable for the fractional data at hand \citep{krisztin2022spatial, lin_2014_binl_v_mnl, papke_1996_401k}. For a comprehensive overview of multinomial logit models, the reader is referred to \citet{agresti_2012_categorical_da}, \citet{cameron_2005_microeconometrics}, and \citet{fox_2015_reg_glm}.

The observations $\bm{Y}_{i t \thickdot}$ can be seen as a sample from a $J$-dimensional multinomial distribution with probabilities $p_{i t j}$, $j = 1, \dots, J$,
\begin{align*}
    \bm{Y}_{i t \thickdot} \overset{\text{ind}}{\sim} \text{Multinomial}(1, {p}_{i t 1}, \dots, {p}_{i t J}),
\end{align*}
forming the kernel of our mixture model. The probabilities are modeled akin to the logistic regression model by linking the covariates, $\bm{X}^{c}$ and $\bm{X}^{nc}$, and regression parameters, $\bm{\beta}$ and $\bm{\theta}$, with the log odds 
\begin{align*}
   &log\left(\frac{p_{i t j}}{\sum_{j = 1}^J p_{i t j}}\right) = \bm{X}^{c, T}_{it \thickdot} \bm{\beta}_{i \thickdot  j} + 
   \bm{X}^{nc, T}_{it \thickdot} \bm{\theta}_{\thickdot  j},
\end{align*}
corresponding to the following representation of the probabilities:
\begin{align}
\label{eq:mnl_prob}
   &p_{i t j} = \frac{exp(\bm{X}^{c, T}_{it \thickdot} \bm{\beta}_{i \thickdot  j} + 
   \bm{X}^{nc, T}_{it \thickdot} \bm{\theta}_{\thickdot  j})}
   {\sum_{j = 1}^J  exp(\bm{X}^{c, T}_{it \thickdot} \bm{\beta}_{i \thickdot  j} + 
   \bm{X}^{nc, T}_{it \thickdot} \bm{\theta}_{\thickdot  j})},
\end{align}
where $\bm{X}^{c, T}_{it \thickdot}$ and $\bm{X}^{nc, T}_{it \thickdot}$ are row vectors containing an intercept and $K$ covariates for unit $i$ at time point $t$ and $\bm{\beta}_{i \thickdot  j}$ and $\bm{\theta}_{\thickdot  j}$ are column vectors of coefficients for areal unit $i$ and category $j$. For identification purposes, it is common practice to designate a baseline category and set the corresponding coefficients to 0. For the remainder of the manuscript, we choose $J$ as the baseline category.  

To complete our model, we specify the base measure $G_0$ and the prior on the homogenous coefficients $F_0$. To stay in a conjugate setting, we use a product of $(J-1) \times K^c$-dimensional and $(J-1) \times K^{nc}$-dimensional normal distributions, respectively.

\section{Bayesian inference}\label{sec:bayes_inf}
In the Bayesian paradigm, closed-form posteriors are not readily available due to the non-linearity of the multinomial logit model. To overcome this, \citet{scott_2011_mlogreg}, \citet{fruehwirth_schnatter_logreg_mix}, and \citet{held_2006_logreg_aux} recast the probability of observing one of $J$ categories into a utility representation \citep{mcfadden_1974_conditional_logit}. In their seminal paper, \citet{polson_2013_logreg_polya} propose a different data augmentation strategy, circumventing the need for the utility representation. The authors show that by conditioning on auxiliary variables following a P\'olya-Gamma (PG) distribution \citep{polson_2013_logreg_polya}, a binary logistic likelihood can be rewritten in Gaussian form. 
This can be exploited to derive the full conditional distributions of $\bm \beta_{i \thickdot j}$ and $\bm \theta_{\thickdot j}$ in a conjugate setting, greatly facilitating efficient sampling. Since we cannot directly sample from the posterior distribution due to its complex form, samples are obtained from the stationary distribution of a suitable Markov chain, formed by the full conditionals of all model parameters. In the following, we outline the MCMC algorithm before presenting the necessary full conditionals.
\sloppy
\paragraph{MCMC algorithm.}
To obtain posterior samples, we start by choosing appropriate starting values, then iteratively cycle through the following steps, and conventionally discard an initial set of draws as burn-in: 
\begin{enumerate}
    \item[1.] For every $i = 1, \dots, N$ sample $ s_i $ from $P(s_i = c | 
        \bm{s}_{-i}, \bm{Y}_{i \thickdot \thickdot}, \bm{X}_{i \thickdot \thickdot},
        \bm{\beta}^{\star},
        \bm{\theta}, \alpha) $ according to Eq.~\eqref{eq:neal_8}.
    \item [2.] For every category $j = 1, \dots J - 1:$ 
    \begin{enumerate}
    \item For every cluster $c = 1, \dots, M:$
        \begin{enumerate}
        \item  Sample $\omega_{s_i = c, t j} \sim 
            p(\omega_{s_i = c, t j} \mid 
            \bm Y_{s_i = c, \thickdot j}, \bm X_{s_i = c, \thickdot \thickdot},
            \bm s, 
            \bm \beta^{\star}_{c \thickdot \thickdot}, \bm \theta)
            $ 
            according to Eq.~\eqref{eq:omega} using those $i$ such that $s_i = c$.
        \item Sample $\bm \beta^{\star}_{c \thickdot j} \sim 
             p(\bm \beta^{\star}_{c \thickdot j} \mid 
            \bm Y_{s_i = c, \thickdot j}, \bm X_{s_i = c, \thickdot \thickdot},
            \bm s, 
            \bm \beta^{\star}_{c \thickdot - j}, \bm \theta_{\thickdot \thickdot },
            \bm{\omega}_{c \thickdot j}) 
            $, according to Eq.~\eqref{eq:beta_post}.
        \end{enumerate}  
        \item Sample $\bm \omega_{i t j}^{L} \sim
            p(\bm \omega_{i t j}^{L} \mid 
            \bm Y_{\thickdot \thickdot j}, \bm X,
            \bm s, 
            \bm \beta^{\star}, \bm \theta)
            $ according to Eq.~\eqref{eq:omega}.
        \item Sample $\bm \theta_{\thickdot j } \sim 
        p(\bm \theta_{ \thickdot j} \mid 
        \bm Y_{\thickdot \thickdot j}, \bm X,
        \bm s, 
        \bm \beta^{\star}, \bm \theta_{\thickdot -j },
        \bm{\omega}_{\thickdot \thickdot j}^{L}) $
        according to Eq.~\eqref{eq:theta_post}.
        \end{enumerate}
    \item [3.] Sample $ \alpha \sim p(\alpha \mid a_{\alpha},b_{\alpha},x)$ using Eq.~\eqref{eq:alpha_post}.
\end{enumerate}

\paragraph{Full conditional of $\bm s$.} Coupling the prior in Eq~\ref{eq:s_prior} with the corresponding likelihood, we can derive the full conditional distribution for $\bm s$ as
\begin{align*}
&P\left(s_i = c \mid s_1, \ldots, s_{i-1}\right) =\\
&\begin{cases} 
\dfrac{n_{ic}}{i-1 + \alpha} 
f\left(\bm{Y}_{s_i = c \thickdot \thickdot} \mid 
\bm{X}^{c}_{s_i = c \thickdot \thickdot}, 
\bm{X}^{nc}_{s_i = c \thickdot \thickdot}, 
\bm{\beta}_{s_i = c \thickdot \thickdot}^{\star}, 
\bm{\theta} \right), & \text{for } c \leq M \\[10pt]
\dfrac{\alpha}{i-1 + \alpha}
\int
f\left(\bm{Y}_{s_i = c \thickdot \thickdot} \mid 
\bm{X}^{c}_{s_i = c \thickdot \thickdot}, 
\bm{X}^{nc}_{s_i = c \thickdot \thickdot}, 
\bm{\beta}_{s_i = c \thickdot \thickdot}^{\star}, 
\bm{\theta} \right)
d G_0, & \text{for } c = M + 1
\end{cases}. \nonumber
\end{align*}

The integral needed to compute the probability of opening a new cluster is only available in closed form in case of a conjugate prior. We resort to the popular Algorithm 8 by \citet{Neal00}, which circumvents the integral by augmenting the state space with a predefined number $N_{aux}$ of potential, auxiliary clusters. Each auxiliary cluster is coupled with a set of auxiliary parameters sampled from the base measure $G_0$, denoted by $\bm{\beta}^{aux} \in \mathbb{R}^{N_{aux} \times K^c \times J}$.
The cluster allocations $\bm s$ can then be assigned to either one of the existing clusters $1,\dots,M$, or one of the auxiliary clusters $M + 1, \dots, M + N_{aux}$ according to the following probabilities: 
\begin{align}\label{eq:neal_8}
       &P(s_i = c | \bm{s}_{-i}, \bm{Y}_{i \thickdot \thickdot}, \bm{X}_{i \thickdot \thickdot},
        \bm{\beta}^{\star},
        \bm{\theta}, \alpha) \\
        & \propto
        \left\{\begin{array}{cl} \frac{n_{-i, c}}{N-1+\alpha} f \left( 
        \bm{Y}_{i \thickdot \thickdot},
        \bm{\beta}^{\star}_{s_i \thickdot \thickdot},
        \bm{\theta}
        \right) & \text { for } 1 \leq c \leq M,\\
 \frac{\alpha / N_{aux}}{N-1+\alpha} 
 f\left(
\bm{Y}_{i \thickdot \thickdot}, \bm{\beta}^{aux}_{s_i \thickdot \thickdot},
        \bm{\theta}
 \right) & \text { for } M < c \leq M + N_{aux}\end{array}\right. \nonumber ,  
\end{align}
where $n_c$ and $n_{-i, c}$ describe the number of units assigned to cluster $c$ with and without unit $i$. Note that while the individual cluster probabilities differ in regard to their $\bm{\beta}$ coefficients, $\bm{\theta}$ contributes equally to all of them.
\paragraph{Full conditional of $\bm \beta$.}
To derive the full conditional distribution for the regression coefficients, we exploit the fact that the conditional likelihood of a multinomial model for a single category $j$ can be rewritten in the form of a binary logistic likelihood \citep{held_2006_logreg_aux} and therefore combined with the auxiliary variable strategy introduced in \citet{polson_2013_logreg_polya}. We start by sampling the auxiliary variables and encode them in a three-dimensional array $ \bm{\omega} \in \mathbb{R}^{N \times T \times J} $
\begin{align}\label{eq:omega}
&\omega_{itj} \sim PG(1, \eta_{i t j}),\\
& \eta_{i t j} =  
\bm{X}_{it\thickdot}^{c,T} \bm{\beta}_{i \thickdot j} +
\bm{X}_{it\thickdot}^{nc,T} \bm{\theta}_{ \thickdot j} - 
C_{i t j}, \nonumber \\
&C_{i t j}= \log \sum_{l \neq j} \exp 
\bm{X}_{it\thickdot}^{c,T} \bm{\beta}_{i \thickdot l} +
\bm{X}_{it\thickdot}^{nc,T} \bm{\theta}_{ \thickdot l}\nonumber.
\end{align}
Conditioning on $\bm{\omega}$, the likelihood for category $j$ can be written as
\begin{align}\label{eq:holmes_cond_lik}
& p( \bm{Y}_{\thickdot \thickdot j}  \mid 
 \bm{\omega},
\bm{X},
\bm{\beta}_{\thickdot \thickdot j}, \bm{\beta}_{\thickdot \thickdot -j},
\bm{\theta}_{ \thickdot j}, \bm{\theta}_{ \thickdot -j}) \propto  
\prod_{t=1}^T \prod_{i=1}^N \exp 
\left\{
\frac{\omega_{itj}}{2}
\left(
\eta_{i t j} -
\kappa_{itj} / \omega_{itj}
\right)^2
\right\}  \nonumber  \\
&\propto \prod_{t=1}^T \exp
\left\{
-\frac{1}{2}
\left(
\bm{X}_{\thickdot t \thickdot}^{c,T} \bm{\beta}_{\thickdot \thickdot j} +
\bm{X}_{\thickdot t \thickdot}^{nc,T} \bm{\theta}_{ \thickdot j} -
C_{\thickdot t j} -
\bm{z}_{tj}
\right)^T 
\bm{\Omega}_{tj} 
(
\bm{X}_{\thickdot t \thickdot}^{c,T} \bm{\beta}_{\thickdot \thickdot j} +
\bm{X}_{\thickdot t \thickdot}^{nc,T} \bm{\theta}_{ \thickdot j} -
C_{\thickdot t j} -
\bm{z}_{tj}
)
\right\}, 
\end{align}
where $\kappa_{itj} = \bm{Y}_{i t j} - \frac{1}{2}$ and 
$\bm{z}_{tj}=\left(\kappa_{1tj} / \omega_{1tj}, \ldots, \kappa_{Ntj} / \omega_{Ntj} \right)$. Furthermore, we denote with $\bm{\Omega}_{tj} = \operatorname{diag}\left(\omega_{1tj}, \ldots, \omega_{Ntj}\right)$ a diagonal matrix. 
We can now compute the joint posterior 
$ p(\bm \beta^{\star}_{c \thickdot j}, \bm \theta_{\thickdot j} \mid 
    \bm Y_{s_i = c, \thickdot j}, \bm X_{s_i = c, \thickdot \thickdot},
    \bm s, 
    \bm \beta^{\star}_{c \thickdot - j}, \bm \theta_{\thickdot -j },
    \bm{\omega}_{s_i = c, \thickdot j}) $  
    by applying Eq.~\eqref{eq:omega} and Eq.~\eqref{eq:holmes_cond_lik} to those $i$ such that $s_i = c$ and combining the likelihood with a normal prior with mean $\bm{\mu}_0 = (\mu_1, \dots, \mu_{J - 1})$ and diagonal covariance matrix 
$\bm{\Sigma}_0 = \operatorname{diag}\left(\sigma^2_{0, 1}, \ldots, \sigma^2_{0, J}\right)$. Due to conjugacy of the normal distribution, the obtained posterior is also normal with covariance matrix and mean given by 
\begin{align*}
 &\overline{\bm{\Sigma}}^c = \left( \sum_{t = 1}^T  \bm{X}_{s_i = c, t \thickdot }^T \bm{\Omega}_{tj}^c \bm{X}_{s_i = c, t \thickdot } + \bm{\Sigma}^{bt,-1}_{ 0} \right)^{-1}, \\
 &\overline{\bm{\mu}}^c =  \overline{\bm{\Sigma}}^c \left(\sum_{t = 1}^T  \bm{X}_{s_i = c, t \thickdot}^T (\bm \kappa_{s_i = c, t j} + \bm{\Omega}_{tj}^c \bm{C}_{s_i = c, t,j}) +  \bm{\Sigma}^{bt,-1}_0 \bm{\mu}^{bt}_{ 0} \right),
\end{align*}
 where $\bm{\Omega}_{tj}^c$ is a diagonal matrix consisting of those $\omega_{itj}$ where $s_i = c$, $\bm{\mu}^{bt}_{0} = \bm{1}_{2} \otimes \bm{\mu}_0$ and 
$\bm{\Sigma}_{0}^{bt} = \bm{I}_{2} \otimes \bm{\Sigma}_0$ with $\bm{1}_{2}$ being a vector of ones of length 2 and $ \bm{I}_{2} $ the identity matrix of dimension $2$. Using standard results for the normal distribution, we can obtain the conditional distribution
$ p(\bm \beta^{\star}_{c \thickdot j} \mid 
    \bm Y_{s_i = c, \thickdot j}, \bm X_{s_i = c, \thickdot \thickdot},
    \bm s, 
    \bm \beta^{\star}_{c \thickdot - j}, \bm \theta_{\thickdot \thickdot},
    \bm{\omega}_{s_i = c, \thickdot j}) $ with posterior mean and covariance matrix 
\begin{align}\label{eq:beta_post}
&\bm{\Sigma}^c_{j} = \overline{\bm{\Sigma}}^c_{c_{\beta}, c_{\beta}} - \overline{\bm{\Sigma}}^c_{c_{\beta}, c_{\theta}} \overline{\bm{\Sigma}}_{c_{\theta}, c_{\theta}}^{c,-1} \overline{\bm{\Sigma}}^c_{c_{\theta}, c_{\beta}}, \\
&\bm{\mu}^c_{j} = \overline{\bm{\mu}}^c_{c_{\beta}} 
+\overline{\bm{\Sigma}}^c_{c_{\beta}, c_{\theta}} \overline{\bm{\Sigma}}_{c_{\theta}, c_{\beta}}^{c,-1}
\left(\bm{\theta}_{\thickdot j}- 
\overline{\bm{\mu}}^c_{c_{\theta}} 
\right), \nonumber
\end{align}
where $c_{\beta} = \{1,\dots,K^c\}$ and $c_{\theta} = \{(K^c + 1),\dots,K\}$ are vectors indexing the cluster- and global-covariates respectively.

\paragraph{Full conditional of $\bm \theta$.}
%
We begin by computing the joint full conditional of all regression coefficients of the $j$th-category, $p(\bm \beta^{\star}_{\thickdot \thickdot j}, \theta_{\thickdot j} \mid 
            \bm Y_{s_i = c, \thickdot j}, \bm X_{s_i = c, \thickdot \thickdot},
            \bm s, 
            \bm \beta^{\star}_{c \thickdot - j}, \bm \theta_{\thickdot -j},
            \bm{\omega}_{s_i = c, \thickdot j}) $. To this end, we construct a new design matrix $ \bm X^{L} \in \mathbb{R}^{N \times T \times (C \times K^{c} + K^{nc})}$ by rearranging parts of $X^{c}$ and $X^{nc}$ that correspond to the individual clusters
 and gather all coefficients by stacking them in the matrix $ \bm{\phi} \in \mathbb{R}^{(C \times K^{c} + K^{nc}) \times J}$.
\begin{align}
   \bm X^{L} = 
  \begin{pmatrix}
  \bm  X^{c}_{s_i = 1 \thickdot \thickdot} & 0 & 0 & 0&\bm X^{nc}_{s_i = 1 \thickdot \thickdot}  \\
  0 &\bm X^{c}_{s_i = 2 \thickdot \thickdot} & 0 & \cdots &\bm X^{nc}_{s_i = 2 \thickdot \thickdot}  \\
  0 & 0 & \ddots & 0 &  \vdots \\
 0 & 0 & 0 &\bm X^{c}_{s_i = M \thickdot \thickdot} &\bm X^{nc}_{s_i = M\thickdot \thickdot}  \\
 \end{pmatrix},
 \quad 
 \bm{\phi} = \begin{bmatrix}
           \bm{\beta}^{\star}_{s_i = 1 \thickdot \thickdot} \\
           \bm{\beta}^{\star}_{s_i = 2 \thickdot \thickdot}  \\
           \vdots \\
           \bm{\theta}_{\thickdot \thickdot} 
\end{bmatrix}.
\end{align}
Applying Eq.~\eqref{eq:omega} and Eq.~\eqref{eq:holmes_cond_lik} to $\bm X^{L}$ and $\bm{\phi}$, we obtain a new set of auxiliary parameters $\bm \omega^L$ and a new likelihood, which we can use to obtain the desired normal posterior distribution with covariance matrix and mean given by
\begin{align*}
         &\overline{\bm{\Sigma}}^{L} = \left( \sum_{t = 1}^T  \bm{X}_{\thickdot t \thickdot }^{L, T} \bm{\Omega}_{tj}^{L} \bm{X}_{\thickdot t \thickdot }^{L} + \bm{\Sigma}^{L,-1}_{0} \right)^{-1}, \\
         &\overline{\bm{\mu}}^{L} =  \overline{\bm{\Sigma}}^L \left(\sum_{t = 1}^T  \bm{X}_{\thickdot t \thickdot }^{L, T} (\bm \kappa_{\thickdot t j} + \bm{\Omega}_{tj}^{L} \bm{C}_{\thickdot t j}^{L}) +  \bm{\Sigma}^{L, -1}_{0} \bm{\mu}_{0}^{L} \right),
\end{align*}
where $\bm{\Omega}_{tj}^{L} = 
\operatorname{diag}\left(\omega_{1tj}^{L}, \ldots, \omega_{Ntj}^{L} \right)$, $\bm{\mu}_{0}^{L} = \bm{1}_{M+1} \otimes \bm{\mu}_0$ and
$\bm{\Sigma}_{0}^{L} = \bm{I}_{(M + 1) } \otimes \bm{\Sigma}_0$.
We can now compute the conditional distribution 
$ p(\bm \theta_{\thickdot j} \mid 
            \bm Y_{s_i = c, \thickdot j}, \bm X_{s_i = c, \thickdot \thickdot},
            \bm s, 
            \bm \beta^{\star}_{c \thickdot \thickdot}, \bm \theta_{\thickdot -j},
            \bm{\Omega}_{c \thickdot j}) $ 
with posterior mean and covariance matrix 
\begin{align}\label{eq:theta_post}
&\bm{\Sigma}_{j} = \overline{\bm{\Sigma}}^{L}_{L_{\theta}, L_{\theta}} - \overline{\bm{\Sigma}}^{L}_{c, L_{\theta}, L_{\beta}} \overline{\bm{\Sigma}}^{L, -1}_{c, L_{\beta}, L_{\beta}} \overline{\bm{\Sigma}}^{L}_{c, L_{\beta},L_{\theta}}, \\
&\bm{\mu}_{j} = \overline{\bm{\mu}}_{c,L_{\theta}} 
+\overline{\bm{\Sigma}}^{L}_{c, L_{\theta}, L_{\beta}} \overline{\bm{\Sigma}}^{L, -1}_{c, L_{\beta}, L_{\beta}}
\left(\bm{\phi}_{\thickdot j}- 
\overline{\bm{\mu}}_{c,L_{\beta}} \nonumber 
\right),
\end{align}
where $L_{\beta} = \{1, \dots, M \times K^{c}\}$ and $L_{\theta} = \{ M \times K^{c} + 1, \dots, M\times K^{c} + K^{nc}\}$.

\paragraph{Full conditional of $\alpha$.}
Adapting the approach in \citet{west_1995_hyperparameter}, who represent the desired posterior of $\alpha$ as a mixture of two gamma densities, we derive the full conditional as
\begin{align}\label{eq:alpha_post}
&\alpha \sim 
\pi_x\Gammad{a_{\alpha}+M, b_{\alpha} - \log(x)}
+
(1-\pi_x)\Gammad{a_{\alpha}+M-1,b_{\alpha}-\log(x)}, \\
&\frac{\pi_x}{1-\pi_x} =\frac{a_{\alpha}+1}{n(b_{\alpha}-log(x))}, \nonumber\\
&x|\alpha,M \sim Beta(\alpha +1 , N), \nonumber
\end{align}
where $x$ is a beta-distributed auxiliary variable.

\section{Data}\label{sec:data_2}

Our dataset encompasses 912 EU regions spread across 21 countries.\footnote{To achieve a balanced panel, we had to discard FADN observations from Bulgaria, Serbia, Poland, Romania, Lithuania, and the United Kingdom.} These regions are categorized according to the NUTS 2016 classification at the NUTS-3 level. \footnote{The Nomenclature of territorial units for statistics, abbreviated NUTS is a geographical nomenclature based on Regulation (EC) No 1059/2003 of the European Parliament and of the Council of 26 May 2003 and subdivides the economic territory of the European Union (EU) into regions at three different levels (NUTS 1, 2 and 3 respectively, ranging from larger to smaller territorial units).}. Although the NUTS-3 regions differ in size, they offer the best possible representation as suitable units for modeling and analysis. It should be noted, however, that NUTS-3 regional boundaries are established based on administrative criteria rather than reflecting functional characteristics. As a result, these boundaries may not correspond precisely to the actual dynamics of the regional processes under study. 

The regions included in the sample are located in Austria (34 regions), Belgium (43 regions), Cyprus (one region), Czech Republic (13 regions), Denmark (9 regions), Finland (18 regions), France (92 regions), Germany (361 regions), Greece (43 regions), Hungary (19 regions), Italy (107 regions), Latvia (10 regions), Lithuania (10 regions), Luxembourg (one region), Malta (one region), Netherlands (38 regions), Portugal (23 regions), Republic of Ireland (eight regions), Slovakia (eight regions), Slovenia (12 regions), Spain (50 regions) and Sweden (21 regions).

The dependent variable in our study is the share of land use in the categories cropland, grassland, forest, urban, and other natural vegetation. Our dataset spans the years 2008 to 2018. The proportions were obtained by applying the methods described in \cite{witjes2022spatiotemporal} to data collected from the Land Use and Coverage Area frame Survey (LUCAS) and the Corine Land Cover (CLC) databases. The land cover data, divided into 33 CLC classes, is consolidated into our five defined land-use categories.

\begin{table}[tp]
  \centering
\begin{threeparttable}
  \caption{Description of dependent and independent variables used in the analysis, including their descriptions and data sources. Data sources include ARDECO (Annual Regional Database of the European Commission's Directorate General for Regional and Urban Policy), CORINE Land Cover, EU-DEM (European Union Digital Elevation Model), and FADN (Farm Accountancy Data Network).}
  \label{tab:variables}
    \begin{tabularx}{\textwidth}{
       >{\RaggedRight\arraybackslash\hsize=.3\hsize\hyphenpenalty=10000\exhyphenpenalty=10000}X
     >{\RaggedRight\arraybackslash\hsize=0.5\hsize}X
      >{\RaggedRight\arraybackslash\hsize=0.2\hsize}X}
    \toprule
    {Variable } & \multicolumn{1}{l}{{Description}} & {Source} \\
    \midrule
    \multicolumn{1}{p{8.665em}}{Cropland} & Sum of cropland cover, divided by \newline{}the total area of the region. &  \\
    \multicolumn{1}{p{8.665em}}{Grassland } & Sum of grassland cover, divided by \newline{}the total area of the region. & \\
    \multicolumn{1}{p{8.665em}}{Forest} & Sum of forest cover, divided by \newline{}the total area of the region. & \citet{parente2021continental} \\
    \multicolumn{1}{p{8.665em}}{Other natural vegetation} & Sum of other natural vegetation cover, divided by \newline{}the total area of the region. &  \\
    \multicolumn{1}{p{8.665em}}{Urban} & Sum of other artifical and urban areas cover, divided by \newline{}the total area of the region. &  \\
    \hline  \\[-1.8ex] 
    Employment primary & Share of employment in the primary sector. & ARDECO \\
    Employment tertiary  & Share of employment in the tertiary sector. & ARDECO \\
    Log GDP per capita & Logarithm of the gross domestic product divided by the population. & ARDECO \\
    Population density & \multicolumn{1}{l}{Population per square kilometer.} & ARDECO \\
    Elevation & \multicolumn{1}{l}{Average elevation in meters.} & EU-DEM \\
    Slope & \multicolumn{1}{l}{Average slope in degree.} & EU-DEM \\
    \hline  \\[-1.8ex] 
    Farm output & Total farm output in EUR divided by utilized agricultural area. &  \\
    Rent  & Total rent paid in EUR divided by the total rented area in ha. &  \\
    Pillar I - Coupled (crops) & Total of Pillar I coupled payments for crops in EUR divided by the total agricultural area in hectare. &  \\
    Pillar I - Coupled (livestock) & Total of Pillar I coupled payments for livestock in EUR divided by the total agricultural area in hectare. & FADN \\
    Pillar I - Decoupled payments & Total of Pillar I decoupled payments  in EUR divided by the total agricultural area in hectare. &  \\
    Pillar II - Environmental payments & Total of Pillar II environmental payments for crops in EUR divided by the total agricultural area in hectare. &  \\
    Pillar II - Least favored area payments & Total of Pillar II least favored area payments for crops in EUR divided by the total agricultural area in hectare. &  \\
    \bottomrule
    \end{tabularx}
  \end{threeparttable}
\end{table}

\begin{table}[tp]
 \caption{Descriptive statistics for the dependent and independent variables used in the analysis, including mean, standard deviation, minimum, $1$st and $3$rd quartile as well as maximum values.}  
  \label{tab:desc_stats}
  \resizebox{\columnwidth}{!}{%
\begin{tabular}{@{\extracolsep{5pt}}lrrrrrrr} 
\toprule \\[-1.8ex] 
Variable & \multicolumn{1}{r}{Mean} & \multicolumn{1}{r}{Std.\ Dev.} & \multicolumn{1}{r}{Min} & \multicolumn{1}{r}{1st Qu.} & \multicolumn{1}{r}{3rd Qu.} & \multicolumn{1}{r}{Max} \\ 
\hline \\[-1.8ex] 
Cropland & 0.32 & 0.19 & 0.00 & 0.17 & 0.46 & 0.81 \\ 
Forest & 0.32 & 0.16 & 0.00 & 0.19 & 0.44 & 0.78 \\ 
Grassland & 0.18 & 0.13 & 0.00 & 0.10 & 0.23 & 0.80 \\ 
Other natural vegetation & 0.09 & 0.11 & 0.00 & 0.02 & 0.13 & 0.53 \\ 
Urban & 0.09 & 0.09 & 0.00 & 0.03 & 0.10 & 0.54 \\ 
\hline \\[-1.8ex] 
Employment primary & 0.05 & 0.06 & 0.00 & 0.02 & 0.06 & 0.48 \\ 
Employment tertiary & 0.76 & 0.09 & 0.46 & 0.71 & 0.83 & 0.96 \\ 
Log GDP per capita & $-$3.66 & 0.42 & $-$5.32 & $-$3.85 & $-$3.42 & $-$2.06 \\ 
Population density & 237.55 & 285.20 & 1.85 & 69.25 & 258.93 & 1,463.02 \\ 
Elevation & 341.59 & 315.23 & $-$3.28 & 99.71 & 477.67 & 2,102.24 \\ 
Slope & 246.56 & 5.20 & 218.00 & 245.73 & 249.77 & 250.00 \\ 
\hline \\[-1.8ex]
Farm output & 4,460.15 & 20,736.51 & 0.00 & 1,723.89 & 3,510.50 & 476,443.70 \\ 
Rent & 264.06 & 339.03 & 0.00 & 129.39 & 324.31 & 8,438.64 \\ 
Coupled (crops) & 12.01 & 22.28 & 0.00 & 0.94 & 14.89 & 221.35 \\ 
Coupled (livestock) & 20.61 & 42.37 & 0.00 & 1.35 & 23.26 & 365.29 \\ 
Decoupled payments & 287.45 & 111.99 & 0.00 & 215.80 & 332.40 & 874.98 \\ 
Environmental payments & 48.95 & 48.91 & 0.00 & 16.82 & 62.22 & 382.30 \\ 
Least favored area payments & 32.65 & 48.53 & 0.00 & 2.14 & 44.90 & 267.05 \\ 
\bottomrule\\[-1.8ex] 
\end{tabular}
}
\end{table}

We utilize a set of variables frequently used in studies on land-use change \citep{meyer2020patterns}, split into $K^c = 7$ cluster-covariates, including an additional intercept term, and $K^{nc} = 6$ global-covariates. Table~\ref{tab:variables} provides overview of our selected variables, while Table~\ref{tab:desc_stats} contains descriptive statistics. The variables are arranged in the tables starting with the dependent variable, followed by the global covariates, and concluding with the cluster-specific covariates.

In order to control for homogenous economic development indicators, we include data on employment, population, and income in our global variables. Employment data are factored in as shares of the respective sector and total employment. Population, a key driver of land use \citep{terama2019modelling}, is adjusted for regional area and thus included as population density. Income levels are represented through the logarithm of the gross domestic product (GDP) per capita.

Additionally, we add physical characteristics of the land, specifically the slope and elevation, to the set of global covariates. The relevance of these features stems from their indirect impact on the costs associated with land-use conversion \citep{chen2023cropland}. Slope and elevation are quantified in average meters and degrees, respectively.

Since the 1960s, the Common Agricultural Policy (CAP) has shaped European agriculture by initially focusing on boosting productivity and ensuring food security through market supports and income subsidies, leading to farming intensification and expansion. Further updates of policies were motivated by environmental concerns, as well as measures addressing rural disparities. This was achieved by splitting policy payments into  direct income support (Pillar I) and rural development measures (Pillar II). The latter includes agri-environmental schemes designed to  promote sustainable practices like organic farming and habitat restoration to enhance biodiversity and reduce environmental impacts.

To analyze these effects, we distinguish between Pillar I and Pillar II payments. Pillar I includes coupled payments (linked to specific commodities, such as the crop and livestock subsidies) and decoupled payments. Pillar II focuses on rural development, including agri-environmental payments and support for Least Favored Areas, following \citet{boulanger2015eu}. 

Economic performance of the agricultural sector is gauged using farm output measured in euros per hectare of utilized agricultural area and land rents in euros per hectare of rented area, sourced from the FADN. Such economic factors are presumed to influence landowners' decisions regarding land use \citep{gorgan2022development} and are therefore included in the list of cluster-covariates.

\section{Results}\label{sec:results}
The following section presents results obtained after applying the BNP-PC model outlined in Section~\ref{sec:multinom} to the dataset described in Section~\ref{sec:data_2}. We begin by detailing the methodology used to generate samples from the posterior distribution and estimate the clustering of the NUTS 3 regions, which forms the foundation for all subsequent analyses and interpretations. This is followed by an in-depth discussion of the effects of CAP subsidies on the three land-use categories: cropland, grassland, and forests, as these are most pertinent to our research objectives. The section concludes with an evaluation of the results concerning the remaining cluster-level covariates, rent and farm output. 

\subsection{Posterior sampling}
\begin{figure}[tp]
    \centering
    \includegraphics[width=0.7\textwidth]{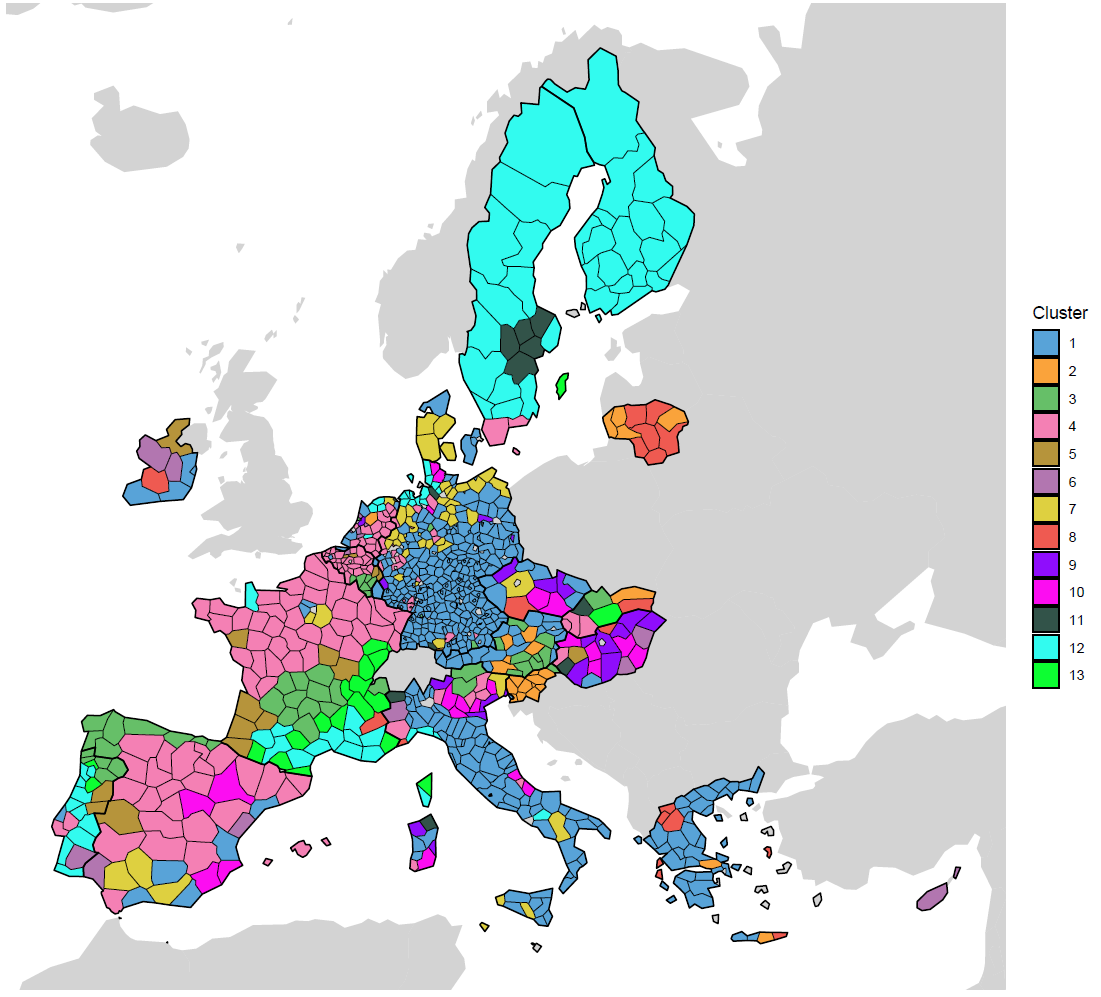}
    \caption{Clustering estimated from posterior samples of the first MCMC run by minimizing the posterior expectation of the Binder loss function.}
    \label{fig:clustering}
\end{figure}

We leverage the MCMC algorithm presented in Section~\ref{sec:bayes_inf} to obtain samples from the posterior of all model parameters, namely the number of clusters $M$, the cluster allocations $\bm s$ and the regression coefficients $\bm \beta$ and $\bm \theta$. We establish convergence of the MCMC chain by inspection of autocorrelation and posterior trace plots. The changing number of clusters and label switching lead to additional challenges in the analysis of the posterior samples. For this reason, we choose to run the sampler once to estimate a representative clustering and a subsequent time, conditional on the estimated clustering, to obtain posterior samples from the remaining parameters that are readily manageable. The estimate of the clustering, depicted in Figure~\ref{fig:clustering}, is based on a sample of 10,000 posterior draws obtained after discarding a burn-in of 5,000 draws. The clustering was estimated by minimizing the posterior expectation of the Binder loss function \citep{binder_1978_binder}, using the stochastic search algorithm from the R package \texttt{salso} \citep{dahl_2022_salso_package}. The second run, used to generate posterior samples obtained for inference on the regression coefficients, consisted of a burn-in and retained sample half the size of the one used to establish the clustering. 

Suitable prior distributions are selected to complete the model specification. For the priors on $\bm{\beta}_{\thickdot \thickdot j}$ and $\bm{\theta}_{\thickdot j}$, $G_0$ and $F_0$, we select hyperparameters that are both fairly general and enable good mixing of the MCMC chain. Specifically, $\bm{\mu}_0$ is a zero vector and $\bm{\Sigma}_0$ a diagonal covariance matrix filled with ones. Regarding the gamma prior for concentration parameter of the DP $\alpha$, we set the hyperparameters to $a_{\alpha} = 3$ and $b_{\alpha} = 2$, implying an interpretable number of clusters $\mathbb{E}[M \mid \alpha] \approx 10$ a priori \citep{teh_2010_dp}. To account for the lag in the effect of the chosen covariates on the corresponding land-use classes, we use $\bm{X}_{i,t-1 \thickdot}$ to estimate $\bm{Y}_{i,t \thickdot}$.


Figure~\ref{fig:clustering} presents the estimated clustering obtained from the first MCMC run. We note significant alignments with biogeographical regions as defined by the European Environment Agency \citep[EEA,][]{eea_2017_biogeo_regions}. Specifically, Clusters 1, 4, and 12 align with the Continental-Mediterranean, Atlantic and Boreal regions, respectively. Clusters with a high share of forest cover, such as clusters 3 and 10, predominantly appear in forested areas of Germany, France, and Scandinavia. This alignment indicates that agricultural practices are closely linked to regional ecological characteristics \citet{rega2019environmentalism}, which must be considered in the application of the agricultural and land use policies. The diversity of European farming systems is reflected in the distinct compositions of the clusters. For instance, Mediterranean clusters (1, 3, and 4) likely adapt to arid conditions, while Boreal clusters (i.e., 12, and 11) reflect more intensive farming suited to cooler, wetter climates. This supports suggestions that CAP interventions should be fine-tuned to regional environmental, economic, and social conditions, as emphasized by \cite{van2015manifestations, piorr2009integrated}, who stress the need for region-specific CAP policies to improve environmental sustainability and economic viability. Moreover, the results show considerable variation within countries, particularly in France, Hungary and The Netherlands, where multiple clusters emerge. 


\begin{table}[tp]
    \centering
    \resizebox{\textwidth}{!}{%
    \begin{threeparttable}
    \begin{tabular}{cp{.6\textwidth}ccccc}
    \toprule
    \multicolumn{1}{l}{Cluster} & \multicolumn{1}{l}{Regions} & Cropland & Grassland & Forest & Other & Urban \\
    \midrule
    1     & DE (278), IT (72), EL (34), AT (20), CZ (6), NL (6), ES (5), DK (4), IE (4), FR (2), BE (1), HU (1), LU (1), PT (1) & 0.346 & 0.156 & 0.331 & 0.071 & 0.096 \\
    2     & SI (11), AT (5), LT (4), DE (2), EL (2), NL (1), SK (1) & 0.135 & 0.233 & 0.468 & 0.128 & 0.036 \\
    3     & FR (12), AT (9), ES (8), BE (7), PT (5), DE (4), IT (2), NL (1), SI (1), SK (1) & 0.115 & 0.301 & 0.446 & 0.084 & 0.054 \\
    4     & FR (39), BE (30), ES (27), DE (21), NL (18), IT (7), SK (3), PT (2), SE (2), DK (1), HU (1) & 0.393 & 0.203 & 0.243 & 0.054 & 0.107 \\
    5     & FR (7), BE (4), PT (2), ES (1), HU (1), IE (1) & 0.209 & 0.304 & 0.301 & 0.117 & 0.069 \\
    6     & NL (5), ES (2), HU (2), IE (2), BE (1), CY (1), IT (1), PT (1) & 0.246 & 0.304 & 0.138 & 0.170 & 0.142 \\
    7     & DE (31), DK (4), IT (4), ES (3), FR (2), CZ (1), NL (1) & 0.446 & 0.165 & 0.254 & 0.048 & 0.087 \\
    8     & EL (7), LT (6), CZ (1), FR (1), HU (1), IE (1), IT (1), SK (1) & 0.244 & 0.212 & 0.290 & 0.232 & 0.022 \\
    9     & HU (7), IT (6), CZ (3), DE (3), NL (1) & 0.389 & 0.159 & 0.257 & 0.105 & 0.090 \\
    10    & IT (8), HU (5), ES (4), DE (3), CZ (2) & 0.503 & 0.126 & 0.185 & 0.111 & 0.075 \\
    11    & SE (5), DE (4), IT (2), HU (1), NL (1), SK (1) & 0.214 & 0.154 & 0.355 & 0.200 & 0.078 \\
    12    & FI (18), DE (15), FR (13), SE (13), PT (10), IT (4), NL (4), MT (1) & 0.192 & 0.173 & 0.358 & 0.217 & 0.060 \\
    13    & FR (16), PT (2), SE (1), SK (1) & 0.147 & 0.243 & 0.422 & 0.141 & 0.047 \\
    \bottomrule
    \end{tabular}
    \end{threeparttable}}
    \caption{Allocation of clusters across countries and land-use types. Numbers in parentheses next to each country code represent the number of NUTS-3 regions from that country included in each cluster. The values in the land-use columns are proportions of the total land-use type within each cluster.}
    \label{tab:cluster_summary}
\end{table}

Table~\ref{tab:cluster_summary} provides an overview of the estimated clusters. The columns of the tables illustrate the share of each cluster per country and across land use types, respectively. For each cluster, the table lists the countries included, with the number of NUTS-3 regions in parentheses, and presents the proportion of different land-use types within each cluster, allowing for easy comparison across regions. The largest clusters are number 1, 4, and 12 in Table~\ref{tab:cluster_summary}. They represent roughly 73 percent of NUTS-3 regions and cover a wide range of land-use types. Cluster 1 predominantly covers Germany, Italy, and Austria, while Cluster 4 is concentrated in northern France, Spain, Belgium, and the Netherlands. Cluster 12 mainly represents Scandinavian countries, such as Finland and Sweden, along with Portugal and parts of France.

\subsection{Impact of common agricultural policy subsidies on land use}

\begin{figure}[t]
    \caption{Top-left: Optimal clustering based on the Binder loss function. Remaining plots: $10\%$, $50\%$ and $90\%$-quantiles of the average posterior marginal effects for the cluster-covariates Pillar I - Coupled (crops), Pillar I - Coupled (livestock), Pillar I - Decoupled, Pillar II - Environmental and Pillar II - Less favored area.}\label{fig:cluster_MFX}
    \includegraphics[width=\textwidth]{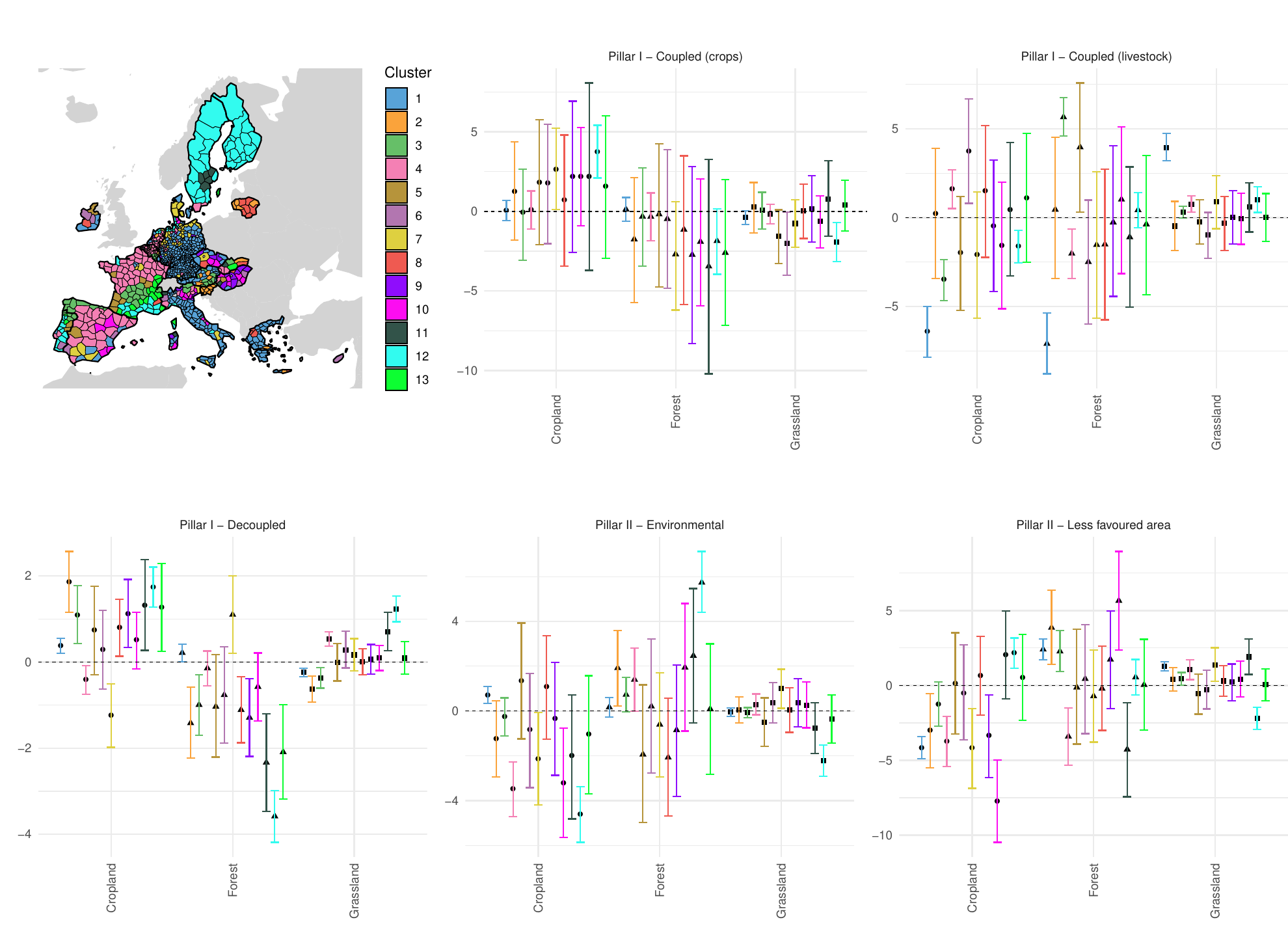}
\end{figure}

The diversity of CAP subsidies, each with distinct and often competing objectives, leads to varied impacts on land use. Several studies \citep[e.g.,][among others]{kirchner2016spatial, brady2012agent, van2006impact} have shown that CAP payments have spatially heterogeneous effects on agricultural land use across regions. To capture these impacts, we examine the effects of five subsidies on three primary land uses: cropland, grassland, and forest. Before we delve into their detailed analysis, Figure~\ref{fig:cluster_MFX} presents an overview of the estimated clustering as well as the average posterior marginal effect of the subsidy-related cluster-covariates on the three land-use categories. In obtaining the marginal effects, we follow the definition of \citet{cameron_2005_microeconometrics}, who define the marginal effect of the $k$th-covariate as the partial derivative of the probability in~\ref{eq:mnl_prob} with respect to this covariate. 

Decoupled payments generally have positive effects across various land types, indicating their flexibility as a policy mechanism without distorting land use decisions. This supports the objectives of the 2013 CAP reforms, which tied greening measures to decoupled payments, promoting environmental sustainability. As noted by \citet{gocht2017eu}, while CAP greening measures have modest overall effects, certain member states experience more pronounced impacts. Similarly, \citet{van2006impact} found that decoupled payments have limited effects on production. In contrast, coupled policies, particularly those for crops and livestock, show mixed impacts, with some negatively affecting forest areas, highlighting potential misalignment with sustainability and conservation targets \citet{brady2012agent}. The variability in these impacts underscores the complexity of agricultural incentives. Pillar II policies, aimed at environmental objectives, generally support forest areas, although their effectiveness varies, particularly in less-favored regions. This suggests a need for better targeting to balance productivity and environmental protection. The average marginal effects presented in Figures~\ref{fig:cropland_mapMFX}–\ref{fig:forest_mapMFX} highlight the impact of ongoing CAP policy refinements on land use, showcasing the need to address the diverse needs of European agriculture. We proceed now with a detailed analysis of the impacts on the three chosen land uses.

\subsubsection{Cropland}

\begin{figure}[tp]
    \caption{Posterior average marginal effect of land-use policies on cropland across Europe.
    The values show a $\%$-change of the underlying area for a change of the subsidy of $100$ EUR per hectare. Grey areas represent effects not significant under a $80\%$ equal-tailed credible interval.}
    \label{fig:cropland_mapMFX}
    \includegraphics[width=\textwidth]{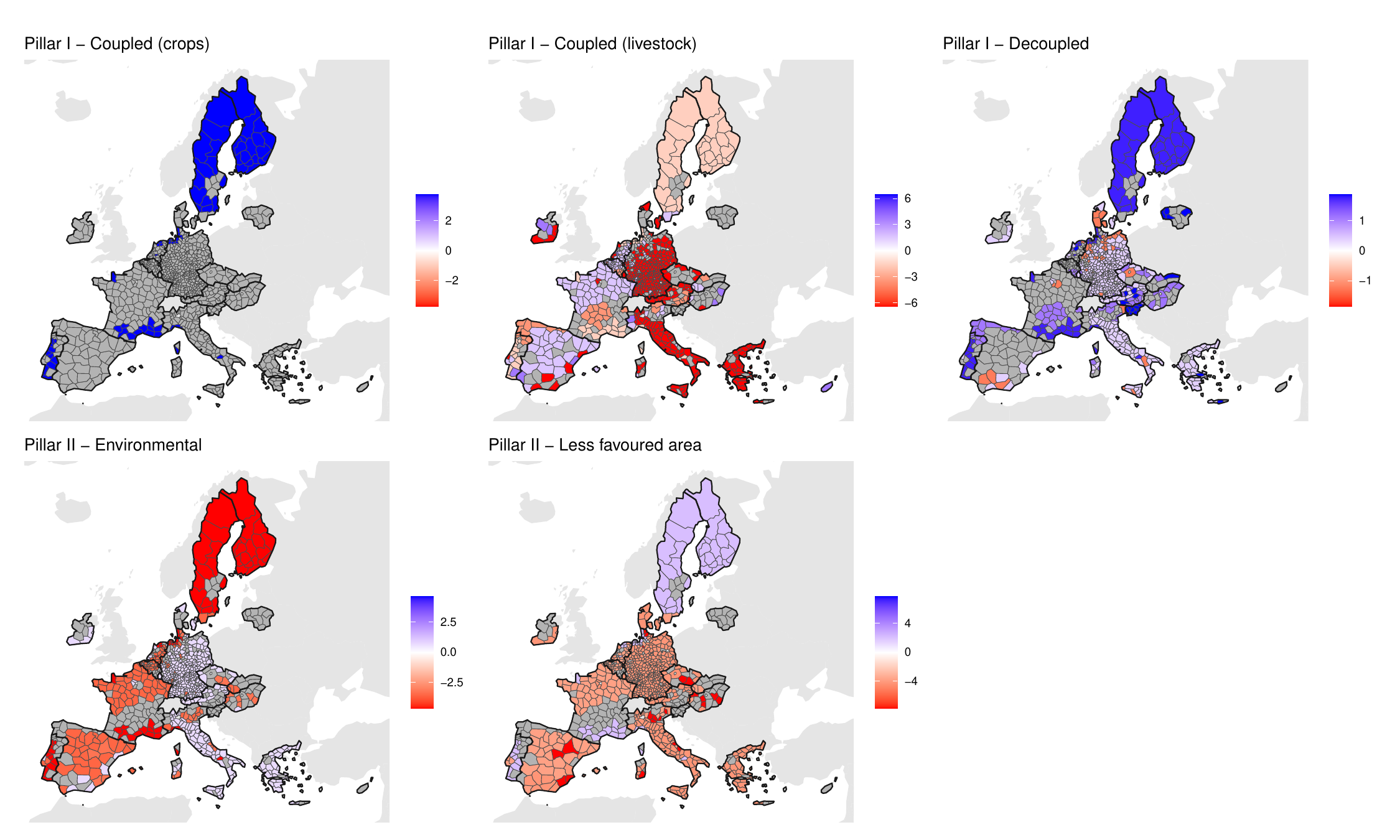}
\end{figure}

As illustrated in Figure~\ref{fig:cropland_mapMFX}, decoupled payments generally exhibit a modest but positive effect on cropland in the EU. A $100$-euro increase per hectare results in a $0.37\%$ rise in cropland share for Cluster 1, while Clusters 2 and 12 experience increases of up to $2\%$. In contrast, Clusters 4 and 7, which already have over $40\%$ cropland coverage, see reductions of $0.4\%$ and $1.2\%$, respectively. This modest impact is consistent with the primary role of decoupled payments as income support, rather than drivers of significant land use changes. Moreover, in regions with high cropland coverage, such as productive agricultural areas, decoupled payments do not incentivize further expansion. These areas show lower sensitivity to subsidy changes, aligning with findings by \citet{helming2011ex}.

Pillar II environmental subsidies reduce cropland shares in Clusters 7, 10, 4, and 12, with decreases ranging from $2\%$  to $4.6\%$. In contrast, Cluster 1 experiences a slight increase of $0.71\%$. The largest reductions are observed in regions characterized by extensive woodland and forest areas, which aligns with the subsidies' objective to promote environmentally sustainable practices and decrease intensive farming. Several conditionalities are associated with this subsidy scheme, supporting the expectation that environmental subsidies tend to encourage less intensive land use and promote transitions to other uses, such as fallow land. This is consistent with the findings of \citet{gocht2017eu}, who report a significant increase in fallow land as farmers either bring new land into cultivation or designate it as ecological focus areas to meet greening requirements.

Similarly, payments for Less Favored Areas (LFA) contribute to reductions in cropland, with declines ranging from $3\%$ in Cluster 2 to $7.7\%$ in Cluster 10 for a $100$-euro increase in subsidies. Although LFA payments aim to prevent land abandonment, they appear insufficient to sustain cropland in less productive regions. However, in Cluster 12, cropland shares increase by $2\%$, suggesting that cropland in more productive northern regions is less sensitive to changes in these payments.

Coupled crop payments positively influence cropland in Clusters 12 and 7, leading to increases of $3.7\%$ and $2.6\%$, respectively. This effect may be attributed to the direct link between the payments and production, which can help sustain or expand cropland, particularly in more productive areas, as suggested by \citet{reger2009potential}. In contrast, the impact of coupled livestock subsidies varies significantly, ranging from a $6.4\%$ reduction in cropland share in Cluster 1 to an $3.7\%$ increase in Cluster 6. This variability indicates that the subsidies can either encourage the expansion of pasture or intensify livestock farming, thus affecting cropland use differently. In regions where crop production supports livestock activities, coupled subsidies may contribute to increases in cropland shares.

\subsubsection{Grasslands}

\begin{figure}[tp]
     \caption{Posterior average marginal effect of land-use policies on grassland across Europe.
    The values show a $\%$-change of the underlying area for a change of the subsidy of $100$ EUR per hectare. Grey areas represent effects not significant under a $80\%$ equal-tailed credible interval.}
    \label{fig:grassland_mapMFX}
    \includegraphics[width=\textwidth]{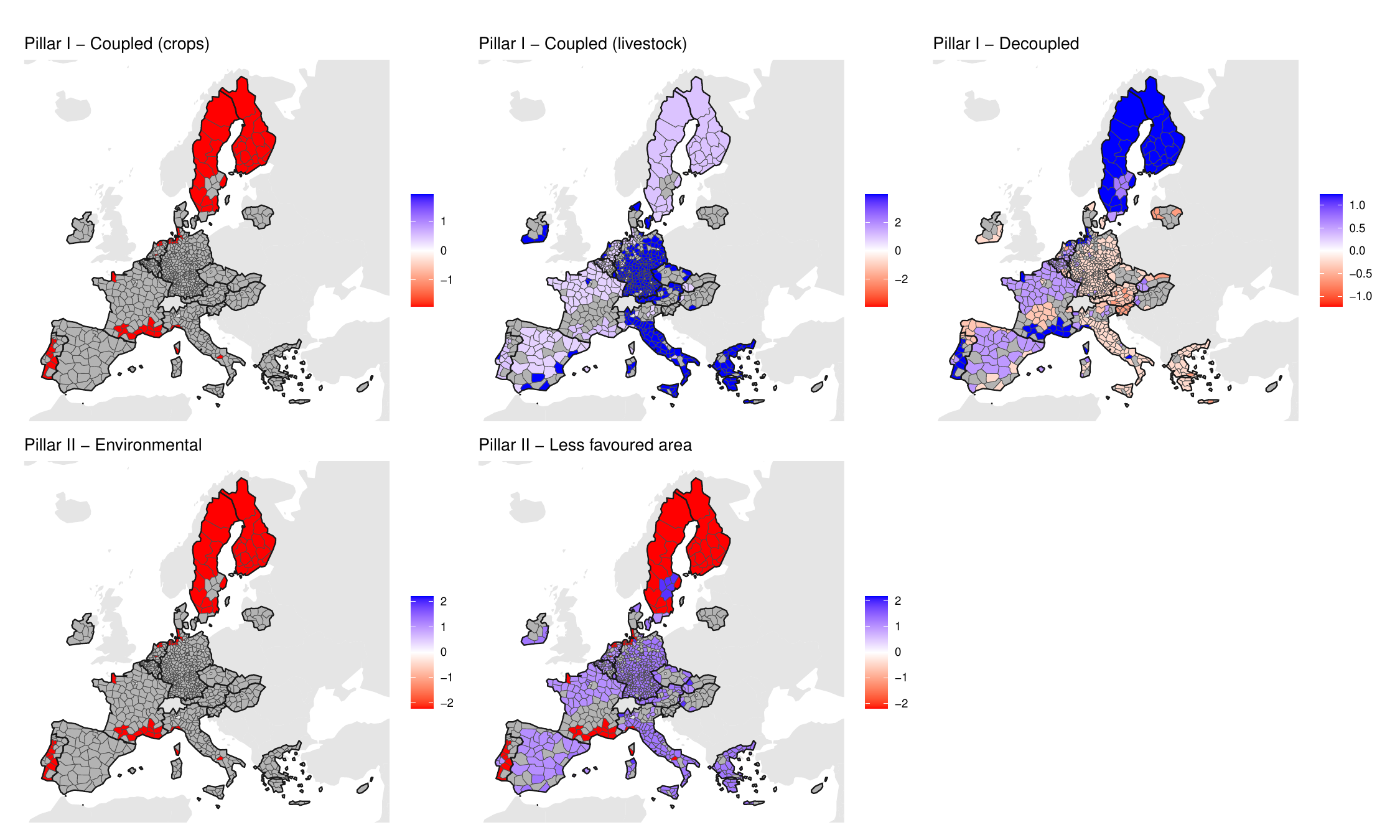}
\end{figure}

The effects of decoupled Pillar 1 payments on grassland (see Figure~\ref{fig:grassland_mapMFX}) are modest and geographically varied, with significant impacts in six clusters. These range from a $0.2\%$ decrease in grassland in Cluster 1 to an $1.2\%$ increase in Cluster 12. These results confirm the minimal effects of decoupled payments on land-use decisions, which can be influenced by economic and environmental factors beyond the subsidies.

Environmental subsidies show more concentrated effects, particularly in Clusters 7 and 12. In Cluster 12, they lead to a $2.2\%$ decrease in grassland, while Cluster 7 sees a $1\%$ increase. This finding is not contradictory but reflect how regional environmental practices and priorities can impact grassland use differently. For example, in regions where environmental practices may conflict with the maintenance or expansion of grasslands, a reduction in grasslands is anticipated. The increase in grassland could suggest that environmental measures may either be compatible with or actively promote grassland use, potentially due to the type of environmental practices encouraged or specific local ecological conditions. LFA payments generally promote grassland, with increases ranging from $0.4\%$ to $2.2\%$, except in Cluster 12, where grassland decreases by $2.2\%$. This could suggest that in more productive areas, LFA payments may not be sufficient to maintain grassland.

Crop subsidies have a modest negative impact on grassland, particularly in Clusters 12 and 6, leading to approximately $2\%$ reductions as land is likely converted to cropland. However, the relatively minor extent reflects the targeted nature of these subsidies. Coupled livestock payments have positive effects on grassland, with increases ranging from $0.76\%$ in Cluster 4 to $4\%$ in Cluster 1. As expected, livestock payment encourage the maintenance or expansion of grassland for grazing. The magnitude of these effects varies by region, reflecting variations in livestock farming intensity and availability of grassland.

\subsubsection{Forests}

\begin{figure}[tp]
    \caption{Posterior average marginal effect of land-use policies on forests across Europe.
    The values show a $\%$-change of the underlying area for a change of the subsidy of $100$ EUR per hectare. Grey areas represent effects not significant under a $80\%$ equal-tailed credible interval.}
    \label{fig:forest_mapMFX}
    \includegraphics[width=\textwidth]{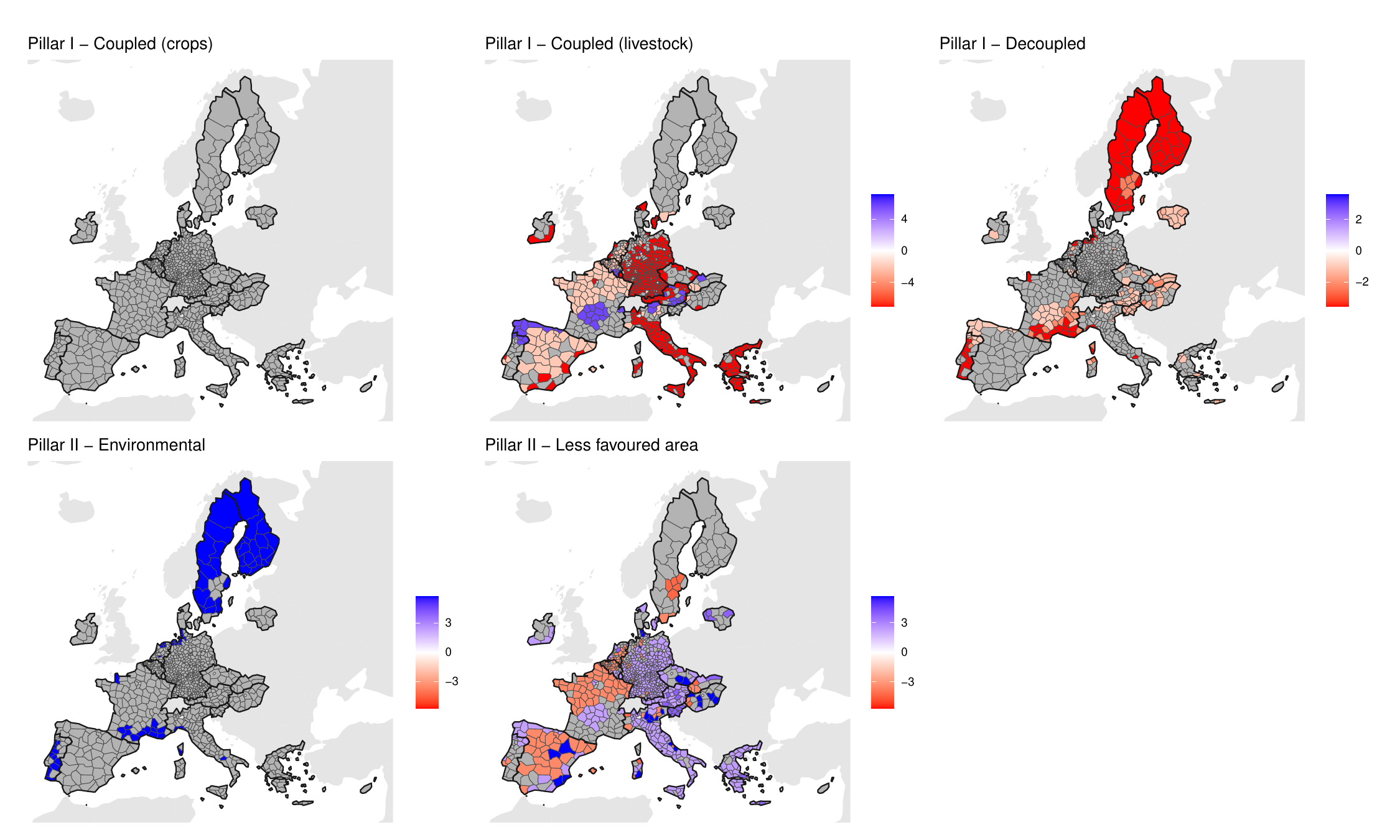}
\end{figure}

The results presented in Figure~\ref{fig:forest_mapMFX} indicate that decoupled payments generally have a negative impact on forest areas across the EU, with decreases ranging from $1\%$ in Cluster 3 to $3.6\%$ in Cluster 12. However, Clusters 7 and 1 show slight increases of $1.1\%$ and $0.2\%$, respectively. Although decoupled payments are not tied to production, they may inadvertently encourage the conversion of forests to other uses, such as cropland or grassland, by providing financial stability for such changes. This inverse relationship between direct payments and forest cover has also been observed by \citet{sieber2013sustainability}.

On the other hand, the slight positive effects in some clusters may reflect regional differences where forest conservation is more economically viable or where other policies have a stronger influence on forest management.

Pillar II environmental subsidies positively affect forest areas, particularly in Clusters 2 and 12. In these clusters, forest cover increases by $1.9 \%$ and $5.7\%$, respectively, with a 100 euros per hectare increase in environmental payments, likely driven by policies that promote landscape conservation. Generally, LFA payments also boost forest cover, with increases ranging from $2.3\%$ in Cluster 3 to $5.7\%$ in Cluster 10. A key objective of LFA payments is to compensate farmers for unfavorable agricultural conditions, thereby discouraging land abandonment. In regions characterized by marginal areas, LFA payments often support land uses that provide additional environmental benefits, which likely explains the overall positive impact on forests.

However, Clusters 11 and 4 experience decreases of $4.3\%$ and $3.4\%$, respectively, suggesting that LFA payments may not always be sufficient to prevent land conversion. Thus, the effectiveness of these payments may depend on local conditions and regional land use pressures.

Crop subsidies show no significant effect on forest areas, suggesting they are not strong enough to drive major changes in forest land use. Coupled livestock payments have mixed effects: forest cover increases by $4\%$ in Cluster 5 and $5.7\%$ in Cluster 3, but decreases by $2\%$ in Cluster 4 and $7 \%$ in Cluster 1. These effects are regionally concentrated, reflecting the varying influence of livestock subsidies on forest dynamics based on local agricultural practices. In some regions, the payments may encourage sustainable livestock farming that coexists with forested areas, while in others, they may drive deforestation to expand pasture land.

The significant negative impacts in certain clusters underscore the potential for these subsidies to conflict with forest conservation goals, especially in areas with intensive livestock production and high agricultural productivity pressures.

\subsection{Impact of land rents and farm output on land use}
The effects of the remaining two cluster-covariates are modest. For cropland, rent shows small impacts (less than $1\%$) in Clusters 1, 2, and 4, with positive effects in Clusters 1 and 2, and negative effects in Cluster 4. Total farm output does not significantly influence cropland shares. For grassland, rent effects are positive but smaller compared to cropland, with increases of approximately $0.2\%$ per 100 euros per hectare observed in Clusters 1 and 12. Overall, land rents have minimal impact on grassland across the EU, and farm output shows no significant effect. Regarding forests, rent has predominantly negative impacts, with forest shares decreasing by $0.5\%$ to $0.9\%$ in Clusters 1, 12, and 2, while slight increases of $0.06\%$ to $0.2\%$ occur in Clusters 3 and 4. Farm output has minimal influence on forest areas, with marginal effects seen only in Clusters 1 and 3.

\section{Conclusion} \label{sec:conclusion_2}

The main contribution of this paper lies in the development of a novel Bayesian nonparametric partial clustering (BNP-PC) framework and its application to assess the heterogeneous effects of agricultural subsidies on land use across the EU. The approach relies on the application of the Dirichlet process as prior for regression coefficients, thereby learning clusters solely from the estimated impact of selected covariates on land use. By further splitting covariates into cluster-specific and global ones, we enable the discovery of clusters based on the impact of agricultural subsidies while controlling for other relevant covariates like GDP or slope and elevation. By incorporating P\'olya-Gamma data augmentation for multinomial logit models, we construct an MCMC algorithm based entirely on full conditional updates, thereby ensuring robust mixing and convergence, eliminating the need for Metropolis-Hastings steps. 

A possible extension would be to vary the set of regression coefficients the DP prior is placed upon. Instead of using it as a prior for a set including all cluster-covariates and categories, interesting alternatives would be to separate DP priors for each category across cluster-covariates or vice versa. These more granular approaches could help address different research questions by uncovering clusters where, for instance, the impact on cropland is consistent across all areas within a cluster, but varies for other land-use categories. Such a model specification would allow for an individual partition for each category, potentially providing a more nuanced and detailed view when needed. 

The analysis reveals that the impact of EU agricultural subsidies on land use is complex and varies significantly across different regions and land types. Decoupled payments, while generally modest in their effects, show a tendency to increase cropland shares in certain clusters while decreasing them in others, particularly in areas with already high cropland shares. Pillar II environmental subsidies demonstrate a strong capacity to reduce cropland shares and promote forest conservation, aligning with their intended goals of encouraging environmentally sustainable practices. Less Favored Areas (LFA) payments generally support the maintenance of grasslands and forests, particularly in regions with natural disadvantages, although their effectiveness appears to vary depending on local conditions. Coupled payments, especially those linked to livestock, show mixed effects on land use, sometimes promoting cropland and grassland expansion at the expense of forests, highlighting the need for careful consideration of these subsidies in the context of broader environmental objectives.

Overall, the results show the need for a detailed approach to design subsidies that account for economic and environmental regional differences. By incorporating insights from the BNP-PC model, policymakers can adjust future policies to insights from the cluster analysis and thereby address regional needs more effectively. 

\printbibliography

@book{agresti_2012_categorical_da,
  title={Categorical Data Analysis},
  author={Agresti, Alan},
  edition={3rd edition},
  year={2012},
  publisher={John Wiley \& Sons},
  DOI={10.1002/0471249688}
}

@article{augustin_2001_spatiotemporal_logit,
 doi = {10.1046/j.1365-2664.2001.00653.x},
 author = {Nicole H. Augustin and Roger P. Cummins and Donald D. French},
 journal = {Journal of Applied Ecology},
 number = {5},
 pages = {991--1006},
 publisher = {[British Ecological Society, Wiley]},
 title = {Exploring Spatial Vegetation Dynamics Using Logistic Regression and a Multinomial Logit Model},
 volume = {38},
 year = {2001}
}

@article{ascolani_2023_clust_consistency,
   author = {F. Ascolani and A. Lijoi and G. Rebaudo and G. Zanella},
   doi = {10.1093/biomet/asac051},
   number = {2},
   journal = {Biometrika},
   title = {Clustering consistency with {D}irichlet process mixtures},
   volume = {110},
   year = {2023},
}

@article{benaglia_2009_mixtools,
   author = {Tatiana Benaglia and Didier Chauveau and David R. Hunter and Derek S. Young},
   doi = {10.18637/jss.v032.i06},
   number = {6},
   journal = {Journal of Statistical Software},
   title = {{mixtools}: An {R} package for analyzing finite mixture models},
   volume = {32},
   year = {2009},
}

@article{binder_1978_binder,
	title={Bayesian cluster analysis},
	author={Binder, David A},
	journal={Biometrika},
	volume={65},
	number={1},
	pages={31--38},
	year={1978},
	publisher={Oxford University Press},
    doi={10.1093/biomet/65.1.31}
}

@article{levers2018spatial,
  title={Spatial variation in determinants of agricultural land abandonment in {E}urope},
  author={Levers, C. and Schneider, M. and Prishchepov, A. and Estel, S. and Kuemmerle, T.},
  journal={The Science of the total environment},
  volume={644},
  pages={95--111},
  year={2018},
  doi={10.1016/j.scitotenv.2018.06.326},
}

@article{blackwell_1973_polya_urn,
author = {David Blackwell and James B. MacQueen},
title = {{F}erguson Distributions Via {P}ólya Urn Schemes},
volume = {1},
journal = {The Annals of Statistics},
number = {2},
publisher = {Institute of Mathematical Statistics},
pages = {353 -- 355},
year = {1973},
doi = {10.1214/aos/1176342372},
}

@article{blei_2006_vi_dpm,
author = {David M. Blei and Michael I. Jordan},
title = {{Variational inference for {D}irichlet process mixtures}},
volume = {1},
journal = {Bayesian Analysis},
number = {1},
publisher = {International Society for Bayesian Analysis},
pages = {121 -- 143},
year = {2006},
doi = {10.1214/06-BA104},
}

@article{boulanger2015eu,
  title={The {EU} budget battle: Assessing the trade and welfare impacts of {CAP} budgetary reform},
  author={Boulanger, Pierre and Philippidis, George},
  journal={Food Policy},
  volume={51},
  pages={119--130},
  year={2015},
  publisher={Elsevier},
  doi = {10.1016/j.foodpol.2015.01.004}
}

@article{burrascano2016current,
  title={Current {E}uropean policies are unlikely to jointly foster carbon sequestration and protect biodiversity},
  author={Burrascano, Sabina and Chytr{\`y}, Milan and Kuemmerle, Tobias and Giarrizzo, Eleonora and Luyssaert, Sebastiaan and Sabatini, Francesco Maria and Blasi, Carlo},
  journal={Biological Conservation},
  volume={201},
  pages={370--376},
  year={2016},
  publisher={Elsevier},
  doi={10.1016/j.biocon.2016.08.005},
}

@book{cameron_2005_microeconometrics,
  title={Microeconometrics: methods and applications},
  author={Cameron, A Colin and Trivedi, Pravin K},
  year={2005},
  publisher={Cambridge university press}
}

@article{chandra_2023_curseodim_mdbclust,
  author  = {Noirrit Kiran Chandra and Antonio Canale and David B. Dunson},
  title   = {Escaping The Curse of Dimensionality in Bayesian Model-Based Clustering},
  journal = {Journal of Machine Learning Research},
  year    = {2023},
  volume  = {24},
  number  = {144},
  pages   = {1--42},
  url     = {https://jmlr.org/papers/v24/21-1276.html}
}

@article{debellagilo_2009_norway_mnl,
title = {Spatial prediction of soil classes using digital terrain analysis and multinomial logistic regression modeling integrated in GIS: Examples from Vestfold County, Norway},
journal = {CATENA},
volume = {77},
number = {1},
pages = {8-18},
year = {2009},
optissn = {0341-8162},
doi = {10.1016/j.catena.2008.12.001},
author = {Misganu Debella-Gilo and Bernd Etzelmüller}
}

@article{west_1995_hyperparameter,
author = {Michael D.   Escobar  and  Mike   West},
title = {Bayesian Density Estimation and Inference Using Mixtures},
journal = {Journal of the American Statistical Association},
pages = {577-588},
volume = {90},
number = {430},
year  = {1995},
publisher = {Taylor & Francis},
doi = {10.1080/01621459.1995.10476550}
}

@article{fraley_1998_howmanyclust,
    author = {Fraley, C. and Raftery, A. E.},
    title = {How Many Clusters? {W}hich Clustering Method?{A}nswers Via Model-Based Cluster Analysis},
    journal = {The Computer Journal},
    volume = {41},
    number = {8},
    pages = {578-588},
    year = {1998},
    doi = {10.1093/comjnl/41.8.578}
}

@incollection{ferguson_1983_dpm,
title = {Bayesian density estimation by mixtures of normal distributions.},
editor = {M. Haseeb Rizvi and Jagdish S. Rustagi and David Siegmund},
booktitle = {Recent Advances in Statistics},
publisher = {Academic Press},
pages = {287-302},
year = {1983},
doi = {10.1016/B978-0-12-589320-6.50018-6},
author = {Thomas S. Ferguson},
}

@article{fruehwirth_schnatter_2019_heretoinfinity,
title = "From here to infinity: {S}parse finite versus {D}irichlet process mixtures in model-based clustering",
author = "Sylvia Fr{\"u}hwirth-Schnatter and Gertraud Malsiner-Walli",
volume={13},
pages = {33--64},
year = "2019",
doi = {10.1007/s11634-018-0329-y},
journal = "Advances in Data Analysis and Classification",
}

@article{fruehwirth_schnatter_logreg_mix,
   author = {Sylvia Frühwirth-Schnatter and Rudolf Frühwirth},
   doi = {10.1016/j.csda.2006.10.006},
   optissn = {01679473},
   number = {7},
   journal = {Computational Statistics and Data Analysis},
   title = {Auxiliary mixture sampling with applications to logistic models},
   volume = {51},
   year = {2007},
pages = {3509-3528}
}

@article{giordano_2023_bnp_sensitivity,
author = {Ryan Giordano and Runjing Liu and Michael I. Jordan and Tamara Broderick},
title = {{Evaluating Sensitivity to the Stick-Breaking Prior in {B}ayesian Nonparametrics (with Discussion)}},
volume = {18},
journal = {Bayesian Analysis},
number = {1},
publisher = {International Society for Bayesian Analysis},
pages = {287 -- 366},
year = {2023},
doi = {10.1214/22-BA1309},
URL = {https://doi.org/10.1214/22-BA1309}
}

@article{gorgan2022development,
  title={Development of agricultural land markets in countries in Eastern {E}urope and Central {A}sia},
  author={Gorgan, Maxim and Hartvigsen, Morten},
  journal={Land Use Policy},
  volume={120},
  pages={106257},
  year={2022},
  publisher={Elsevier},
doi={10.1016/j.landusepol.2022.106257}

}

@article{gruen_2008_mulnom_mix,
   author = {Bettina Grün and Friedrich Leisch},
   doi = {10.1007/s00357-008-9022-8},
   optissn = {14321343},
   number = {2},
   pages = {225--247},
   journal = {Journal of Classification},
   title = {Identifiability of finite mixtures of multinomial logit models with varying and fixed effects},
   volume = {25},
   year = {2008},
}

@misc{eea_2017_biogeo_regions,
  author       = {{European Environment Agency}},
  title        = {Biogeographical Regions in {E}urope},
  year         = {2016},
  publisher    = {{European Environment Agency}},
  url          = {https://www.eea.europa.eu/data-and-maps/data/biogeographical-regions-europe-3}
}

@article{chen2023cropland,
  title={Cropland carbon stocks driven by soil characteristics, rainfall and elevation},
  author={Fangzheng Chen and Puyu Feng and Matthew Tom Harrison and Bin Wang and Ke Liu and Chenxia Zhang and Kelin Hu},
  journal={Science of The Total Environment},
  volume={862},
  pages={160602},
  year={2023},
  publisher={Elsevier},
  doi = {10.1016/j.scitotenv.2022.160602}
}

@article{giannakis2015highly,
  title={The highly variable economic performance of {E}uropean agriculture},
  author={Giannakis, Elias and Bruggeman, Adriana},
  journal={Land Use Policy},
  volume={45},
  pages={26--35},
  year={2015},
  publisher={Elsevier},
  doi = {10.1016/j.landusepol.2014.12.009},
}

@article{helming2018economic,
  title={The economic, environmental and agricultural land use effects in the {E}uropean Union of agricultural labour subsidies under the Common Agricultural Policy},
  author={Helming, J. and Tabeau, A.},
  journal={Regional Environmental Change},
  volume={18},
  number={3},
  pages={763--773},
  year={2018},
  publisher={Springer},
  doi={10.1007/s10113-016-1095-z},
}

@article{hao2015integration,
  title={Integration of multinomial-logistic and {M}arkov-chain models to derive land-use change dynamics},
  author={Hao, Cui and Zhang, Jiahua and Li, Hongyuan and Yao, Fengmei and Huang, Huanchun and Meng, Weiqing},
  journal={Journal of Urban Planning and Development},
  volume={141},
  number={3},
  pages={05014017},
  year={2015},
  publisher={American Society of Civil Engineers},
DOI={10.1061/(ASCE)UP.1943-5444.0000222}

}

@article{hulber2017plant,
  title={Plant species richness decreased in semi-natural grasslands in the Biosphere Reserve {W}ienerwald, {A}ustria, over the past two decades, despite agri-environmental measures},
  author={Hülber, K. and Moser, D. and Sauberer, N. and Maas, B. and Staudinger, M. and Grass, V. and Wrbka, T. and Willner, W.},
  journal={Agriculture, Ecosystems \& Environment},
  volume={243},
  pages={10-18},
  year={2017},
  doi={10.1016/J.AGEE.2017.04.002},
}

@article{oneill2020forest,
doi = {10.1088/1748-9326/abaf87},
opturl = {https://dx.doi.org/10.1088/1748-9326/abaf87},
year = {2020},
optmonth = {oct},
publisher = {IOP Publishing},
volume = {15},
number = {10},
pages = {104090},
author = {Connie O’Neill and Felix K S Lim and David P Edwards and Colin P Osborne},
title = {Forest regeneration on {E}uropean sheep pasture is an economically viable climate change mitigation strategy},
journal = {Environmental Research Letters},
}

@article{rega2019environmentalism,
title = {Environmentalism and localism in agricultural and land-use policies can maintain food production while supporting biodiversity. Findings from simulations of contrasting scenarios in the {EU}},
journal = {Land Use Policy},
volume = {87},
pages = {103986},
year = {2019},
optissn = {0264-8377},
doi = {10.1016/j.landusepol.2019.05.005},
opturl = {https://www.sciencedirect.com/science/article/pii/S0264837718315631},
author = {Carlo Rega and John Helming and Maria Luisa Paracchini}
}

@article{plieninger2016driving,
  title={The driving forces of landscape change in {E}urope: A systematic review of the evidence},
  author={Plieninger, Tobias and Draux, H{\'e}l{\`e}ne and Fagerholm, Nora and Bieling, Claudia and B{\"u}rgi, Matthias and Kizos, Thanasis and Kuemmerle, Tobias and Primdahl, J{\o}rgen and Verburg, Peter H},
  journal={Land Use Policy},
  volume={57},
  pages={204--214},
  year={2016},
  publisher={Elsevier},
  doi={10.1016/j.landusepol.2016.04.040}
}

@article{peer2019greener,
  title={A greener path for the {EU} Common Agricultural Policy},
  author={Pe'er, G. and Zinngrebe, Y. and Moreira, F. and Sirami, C. and Schindler, S. and Müller, R. and Bontzorlos, V. and Clough, D. and Bezák, P. and Bonn, A. and others},
  journal={Science},
  volume={365},
  number={6452},
  pages={449--451},
  year={2019},
  publisher={American Association for the Advancement of Science},
  doi={10.1126/science.aax3146},
}

@article{terama2019modelling,
  title={Modelling population structure in the context of urban land use change in {E}urope},
  author={Terama, Emma and Clarke, Elizabeth and Rounsevell, Mark DA and Fronzek, Stefan and Carter, Timothy R},
  journal={Regional environmental change},
  volume={19},
  number={3},
  pages={667--677},
  year={2019},
  publisher={Springer},
DOI={10.1007/s10113-017-1194-5}
}

@article{parente2021continental,
  title        = {Continental {E}urope land cover mapping at 30m resolution based {{CORINE}} and {LUCAS} on samples},
  author       = {Parente, Leandro and Witjes, Martijn and Hengl, Tomislav and Landa, Martin and Brodsky, Lukas},
  year         = {2021},
  version      = {v0.1},
  publisher    = {Zenodo},
  doi          = {10.5281/zenodo.4725429},
}

@article{witjes2022spatiotemporal,
  title={A spatiotemporal ensemble machine learning framework for generating land use/land cover time-series maps for {E}urope (2000--2019) based on {LUCAS}, {CORINE} and {GLAD} Landsat},
  author={Witjes, Martijn and Parente, Leandro and van Diemen, Chris J and Hengl, Tomislav and Landa, Martin and Brodsk{\`y}, Luk{\'a}{\v{s}} and Halounova, Lena and Kri{\v{z}}an, Josip and Antoni{\'c}, Luka and Ilie, Codrina Maria and others},
  journal={PeerJ},
  volume={10},
  pages={e13573},
  year={2022},
  publisher={PeerJ Inc.},
 doi= {10.7717/peerj.13573}
}

@article{meyer2020patterns,
  title={Patterns and drivers of recent agricultural land-use change in {S}outhern {G}ermany},
  author={Meyer, Markus A and Fr{\"u}h-M{\"u}ller, Andrea},
  journal={Land Use Policy},
  volume={99},
  pages={104959},
  year={2020},
  publisher={Elsevier},
doi={10.1016/j.landusepol.2020.104959}
}

@article{temme2011mapping,
  title={Mapping and modelling of changes in agricultural intensity in {E}urope},
  author={Temme, AJAM and Verburg, PH},
  journal={Agriculture, Ecosystems \& Environment},
  volume={140},
  number={1},
  pages={46--56},
  year={2011},
  publisher={Elsevier},
  doi ={10.1016/j.agee.2010.11.010}
}

@article{krisztin2022spatial,
  title={A spatial multinomial logit model for analysing urban expansion},
  author={Krisztin, Tam{\'a}s and Piribauer, Philipp and W{\"o}gerer, Michael},
  journal={Spatial Economic Analysis},
  volume={17},
  number={2},
  pages={223--244},
  year={2022},
  publisher={Taylor \& Francis},
doi={10.1080/17421772.2021.1933579}

}

@article{held_2006_logreg_aux,
author = {Leonhard Held and Chris C. Holmes},
title = {Bayesian auxiliary variable models for binary and multinomial regression},
volume = {1},
journal = {Bayesian Analysis},
number = {1},
publisher = {International Society for Bayesian Analysis},
pages = {145 -- 168},
year = {2006},
doi = {10.1214/06-BA105},
}

@book{fox_2015_reg_glm,
  title={Applied regression analysis and generalized linear models},
  author={Fox, John},
  year={2015},
  publisher={Sage publications}
}

@article{lin_2014_binl_v_mnl,
  title    = "Comparison of multinomial logistic regression and logistic regression: which is more efficient in allocating land use?",
  author   = "Lin, Yingzhi and Deng, Xiangzheng and Li, Xing and Ma, Enjun",
  journal  = "Frontiers of Earth Science",
  volume   =  8,
  number   =  4,
  pages    = "512--523",
  year     =  2014,
doi = {10.1007/s11707-014-0426-y}
}

@article{Lo_1984_dpms,
 author = {Albert Y. Lo},
 journal = {The Annals of Statistics},
 number = {1},
 pages = {351--357},
 publisher = {Institute of Mathematical Statistics},
 title = {On a Class of {B}ayesian Nonparametric Estimates: I. Density Estimates},
 volume = {12},
 year = {1984},
DOI= {10.1214/aos/1176346412}
}

@article{maceachern_1998_nogaps,
 author = {Steven N. MacEachern and Peter Müller},
 journal = {Journal of Computational and Graphical Statistics},
 number = {2},
 pages = {223--238},
 publisher = {[American Statistical Association, Taylor & Francis, Ltd., Institute of Mathematical Statistics, Interface Foundation of America]},
 title = {Estimating Mixture of {D}irichlet Process Models},
 volume = {7},
 year = {1998},
doi={10.2307/1390815}
}

@incollection{mcfadden_1974_conditional_logit,
	optaddress = {New York},
	title = {Conditional logit analysis of qualitative choice behavior},
	booktitle = {Fontiers in {Econometrics}},
	publisher = {Academic press},
	author = {McFadden, Daniel},
	editor = {Zarembka, Paul},
	year = {1974},
	pages = {105--142}
}

@article{miller_2014_inconsistency,
  title={Inconsistency of {P}itman-{Y}or process mixtures for the number of components},
  author={Miller, Jeffrey W and Harrison, Matthew T},
  journal={The Journal of Machine Learning Research},
  volume={15},
  number={1},
  pages={3333--3370},
  year={2014},
}

@article{mozdzen_2022_bstc,
title = {Bayesian modeling and clustering for spatio-temporal areal data: An application to {I}talian unemployment},
journal = {Spatial Statistics},
volume = {52},
pages = {100715},
year = {2022},
optissn = {2211-6753},
doi = {10.1016/j.spasta.2022.100715},
author = {Alexander Mozdzen and Andrea Cremaschi and Annalisa Cadonna and Alessandra Guglielmi and Gregor Kastner},
}

@article{cortignani2019cap,
  title={{CAP}'s environmental policy and land use in arable farms: An impacts assessment of greening practices changes in {I}taly},
  author={Cortignani, Raffaele and Dono, Gabriele},
  journal={Science of the Total Environment},
  volume={647},
  pages={516--524},
  year={2019},
  publisher={Elsevier},
 doi = {10.1016/j.scitotenv.2018.07.443}
}

@article{papastamoulis_2023_model_clustering,
author={Papastamoulis, Panagiotis},
title={Model based clustering of multinomial count data},
journal={Advances in Data Analysis and Classification},
year={2023},
day={05},
optissn={1862-5355},
doi={10.1007/s11634-023-00547-5},
}

@article{papke_1996_401k,
author = {Papke, Leslie E. and Wooldridge, Jeffrey M.},
title = {Econometric methods for fractional response variables with an application to 401(k) plan participation rates},
journal = {Journal of Applied Econometrics},
volume = {11},
number = {6},
pages = {619-632},
doi = {10.1002/(SICI)1099-1255(199611)11:6<619::AID-JAE418>3.0.CO;2-1},
year = {1996}
}

@article{polson_2013_logreg_polya,
   author = {Nicholas G. Polson and James G. Scott and Jesse Windle},
   doi = {10.1080/01621459.2013.829001},
   optissn = {1537274X},
   number = {504},
   journal = {Journal of the American Statistical Association},
   title = {Bayesian inference for logistic models using {P}ólya-Gamma latent variables},
   volume = {108},
   year = {2013},
}

@article{dahl_2022_salso_paper,
author = {David B. Dahl and Devin J. Johnson and Peter Müller},
title = {Search Algorithms and Loss Functions for Bayesian Clustering},
journal = {Journal of Computational and Graphical Statistics},
volume = {31},
number = {4},
pages = {1189--1201},
year = {2022},
publisher = {Taylor \& Francis},
doi = {10.1080/10618600.2022.2069779},
}

@Manual{dahl_2022_salso_package,
    title = {salso: Search Algorithms and Loss Functions for Bayesian Clustering},
    author = {David B. Dahl and Devin J. Johnson and Peter Müller},
    year = {2023},
    note = {R package version 0.3.35},
    url = {https://CRAN.R-project.org/package=salso}
}

@article{scott_2011_mlogreg,
  title={Data augmentation, frequentist estimation, and the {B}ayesian analysis of multinomial logit models},
  author={Steven L Scott},
  journal={Statistical Papers},
  volume={52},
  number={1},
  pages={87--109},
  year={2011},
  publisher={Springer},
  doi = {10.1007/s00362-009-0205-0}
}

@Inbook{teh_2010_dp,
author={Teh, Yee Whye},
title={Dirichlet Process},
bookTitle="Encyclopedia of Machine Learning",
year={2010},
publisher={Springer US},
optaddress="Boston, MA",
pages={280--287},
optisbn="978-0-387-30164-8",
doi={10.1007/978-0-387-30164-8\_219}
}

@article{van2015manifestations,
  title={Manifestations and underlying drivers of agricultural land use change in {E}urope},
  author={Van Vliet, Jasper and de Groot, Henri LF and Rietveld, Piet and Verburg, Peter H},
  journal={Landscape and Urban Planning},
  volume={133},
  pages={24--36},
  year={2015},
  publisher={Elsevier},
  doi = {10.1016/j.landurbplan.2014.09.001}
}

@article{van2006impact,
title = {The impact of different policy environments on agricultural land use in {E}urope},
journal = {Agriculture, Ecosystems \& Environment},
volume = {114},
number = {1},
pages = {21-38},
year = {2006},
doi = {10.1016/j.agee.2005.11.006},
author = {H. {van Meijl} and T. {van Rheenen} and A. Tabeau and B. Eickhout},
}

@article{renwick2013policy,
  title={Policy reform and agricultural land abandonment in the {EU}},
  author={Renwick, Alan and Jansson, Torbjorn and Verburg, Peter H and Revoredo-Giha, Cesar and Britz, Wolfgang and Gocht, Alexander and McCracken, Davy},
  journal={Land use policy},
  volume={30},
  number={1},
  pages={446--457},
  year={2013},
  publisher={Elsevier},
  DOI = {10.22004/ag.econ.108772}
}

@article{kuemmerle2016hotspots,
  title={Hotspots of land use change in {E}urope},
  author={Kuemmerle, Tobias and Levers, Christian and Erb, Karlheinz and Estel, Stephan and Jepsen, Martin R and M{\"u}ller, Daniel and Plutzar, Christoph and St{\"u}rck, Julia and Verkerk, Pieter J and Verburg, Peter H and others},
  journal={Environmental research letters},
  volume={11},
  number={6},
  pages={064020},
  year={2016},
  publisher={IOP Publishing},
DOI = {10.1088/1748-9326/11/6/064020}

}

@article{reidsma2006impacts,
  title={Impacts of land-use change on biodiversity: an assessment of agricultural biodiversity in the {E}uropean {U}nion},
  author={Reidsma, Pytrik and Tekelenburg, Tonnie and Van den Berg, Maurits and Alkemade, Rob},
  journal={Agriculture, ecosystems \& environment},
  volume={114},
  number={1},
  pages={86--102},
  year={2006},
  publisher={Elsevier},
doi = {10.1007/s00267-008-9270-8}

}

@article{overmars2013modelling,
  title={A modelling approach for the assessment of the effects of Common Agricultural Policy measures on farmland biodiversity in the {EU27}},
  author={Overmars, Koen P and Helming, John and van Zeijts, Henk and Jansson, Torbj{\"o}rn and Terluin, Ida},
  journal={Journal of environmental management},
  volume={126},
  pages={132--141},
  year={2013},
  publisher={Elsevier},
doi ={10.1016/j.jenvman.2013.04.008}

}

@article{piorr2009integrated,
  title={Integrated assessment of future {CAP} policies: land use changes, spatial patterns and targeting},
  author={Piorr, Annette and Ungaro, Fabrizio and Ciancaglini, Arianna and Happe, Kathrin and Sahrbacher, Amanda and Sattler, Claudia and Uthes, Sandra and Zander, Peter},
  journal={environmental science \& policy},
  volume={12},
  number={8},
  pages={1122--1136},
  year={2009},
  publisher={Elsevier},
doi={10.1016/j.envsci.2009.01.001}

}

@article{kirchner2016spatial,
  title={Spatial impacts of the {CAP} post-2013 and climate change scenarios on agricultural intensification and environment in {A}ustria},
  author={Kirchner, Mathias and Sch{\"o}nhart, Martin and Schmid, Erwin},
  journal={Ecological Economics},
  volume={123},
  pages={35--56},
  year={2016},
  publisher={Elsevier},
doi={10.1016/j.ecolecon.2015.12.009}
}

@article{brady2012agent,
  author       = {Mark Brady and Christoph Sahrbacher and Konrad Kellermann and Kathrin Happe},
  title        = {An agent-based approach to modeling impacts of agricultural policy on land use, biodiversity and ecosystem services},
  journal      = {Landscape Ecology},
  year         = {2012},
  volume       = {27},
  number       = {9},
  pages        = {1363--1381},
  doi          = {10.1007/s10980-012-9787-3},
  opturl          = {https://doi.org/10.1007/s10980-012-9787-3},
  optissn         = {1572-9761}
}

@article{gocht2017eu,
  title={{EU}-wide economic and environmental impacts of {CAP} greening with high spatial and farm-type detail},
  author={Gocht, Alexander and Ciaian, Pavel and Bielza, Maria and Terres, Jean-Michel and R{\"o}der, Norbert and Himics, Mihaly and Salputra, Guna},
  journal={Journal of Agricultural Economics},
  volume={68},
  number={3},
  pages={651--681},
  year={2017},
  publisher={Wiley Online Library},
 doi={10.1111/1477-9552.12217}
}

@article{helming2011ex,
  title={Ex ante impact assessment of policies affecting land use, Part A: analytical framework},
  author={Helming, Katharina and Diehl, Katharina and Bach, Hanne and Dilly, Oliver and K{\"o}nig, Bettina and Kuhlman, Tom and P{\'e}rez-Soba, Marta and Sieber, Stefan and Tabbush, Paul and Tscherning, Karen and others},
  journal={Ecology and Society},
  volume={16},
  number={1},
  pages={29},
  year={2011},
  publisher={JSTOR},
DOI = {10.5751/ES-03840-160129}
}

@article{reger2009potential,
  title={Potential effects of direct transfer payments on farmland habitat diversity in a marginal {E}uropean landscape},
  author={Reger, Birgit and Sheridan, Patrick and Simmering, Dietmar and Otte, Annette and Waldhardt, Rainer},
  journal={Environmental management},
  volume={43},
  number={6},
  pages={1026--1038},
  year={2009},
  publisher={Springer},
  doi={10.1007/s00267-008-9270-8}
}

@article{sieber2013sustainability,
title = {Sustainability impact assessment using integrated meta-modelling: Simulating the reduction of direct support under the {EU} common agricultural policy {(CAP)}},
journal = {Land Use Policy},
volume = {33},
pages = {235-245},
year = {2013},
optissn = {0264-8377},
doi = {10.1016/j.landusepol.2013.01.002},
opturl = {https://www.sciencedirect.com/science/article/pii/S0264837713000215},
author = {S. Sieber and T.S. Amjath-Babu and T. Jansson and K. Müller and K. Tscherning and F. Graef and D. Pohle and K. Helming and B. Rudloff and B.S. Saravia-Matus and S. {Gomez y Paloma}}
}

@article{wade_2023_bayclust_analysis,
author = {Wade, Sara},
title = {Bayesian cluster analysis},
journal = {Philosophical Transactions of the Royal Society A: Mathematical, Physical and Engineering Sciences},
volume = {381},
number = {2247},
pages = {20220149},
year = {2023},
doi = {10.1098/rsta.2022.0149},
}

@article{vi,
	title={Comparing clusterings---an information based distance.},
	author={Meil{\u{a}}, Marina},
	journal={Journal of Multivariate Analysis},
	volume={98},
	number={5},
	pages={873--895},
	year={2007},
	publisher={Elsevier},
	doi={10.1016/j.jmva.2006.11.013}
}

@article{mfm,
author = {Jeffrey W Miller and Matthew T Harrison},
title = {Mixture Models With a Prior on the Number of Components},
journal = {Journal of the American Statistical Association},
volume = {113},
number = {521},
pages = {340-356},
year  = {2018},
publisher = {Taylor \& Francis},
doi = {10.1080/01621459.2016.1255636},
}

@article{Neal00,
 author = {Radford M. Neal},
 journal = {Journal of Computational and Graphical Statistics},
 number = {2},
 pages = {249--265},
 title = {Markov Chain Sampling Methods for {D}irichlet Process Mixture Models},
 volume = {9},
 year = {2000},
 doi = {10.2307/1390653},
}

@article {sethuraman_94_stick_break,
	AUTHOR = {Sethuraman, J.},
	TITLE = {A constructive definition of {D}irichlet priors},
	JOURNAL = {Statistica Sinica},
	VOLUME = {4},
	YEAR = {1994},
	NUMBER = {2},
	PAGES = {639--650},
	url = {http://www.jstor.org/stable/24305538}
}

\end{document}